\documentclass[11pt,british]{article}
\usepackage[T1]{fontenc}
\usepackage[utf8]{inputenc}
\usepackage{geometry}
\usepackage{booktabs,siunitx}
\usepackage{textcomp}
\usepackage{amstext}
\usepackage{setspace}
\usepackage[authoryear]{natbib}
\usepackage{hyperref}
\makeatletter

\usepackage{authblk}
\usepackage{adjustbox}
\usepackage{graphicx}
\usepackage{array}
\usepackage{rotating}
\usepackage{pdflscape}
\renewenvironment{sidewaystable}[1][]{\begin{landscape}\begin{table}[!htbp]}{\end{table}\end{landscape}}
\usepackage{float}
\usepackage{booktabs}
\usepackage{stackrel}
\usepackage[skip=1pt, font={sc}]{caption}
\usepackage{varwidth}
\usepackage{subcaption}
\usepackage{bbding}

\usepackage{amsmath}
\usepackage{amsfonts}

\usepackage{tikz}

\usepackage{placeins}

\usepackage{xcolor}

\defcitealias{CORPAUTH}{RLCAN}

\makeatother

\usepackage{babel}

\usepackage{array}
\newcolumntype{H}{>{\setbox0=\hbox\bgroup}c<{\egroup}@{}}

\begin{document}
















\title{\textbf{Work that Counts, and Work that Doesn't: \\ Reconciling Experimental and Observational Labor Supply Elasticities}\thanks{An early version of this paper was circulated with the title "Count Your Losses, and Cut Your Blessings: Reference Dependence across Intertemporal and Uncompensated Labor Supply". The authors wish to thank Davide Marchiori for his work on the real-effort experiment and his feedback on the whole project, and Ozge Demirci and Alessandro Pizzigolotto for their advice and suggestions. The authors would also like to thank the participants of the 2026 EAYE Annual Meeting and the 2026 Royal Economic Society Annual Conference for their precious comments. All errors remain ours. This research was supported by the “Resilienza Economica e Digitale” project (CUP D67G23000060001) funded by the Italian Ministry of University and Research (MUR) as “Department of Excellence” (Dipartimenti di Eccellenza 2023-2027, Ministerial Decree no. 230/2022).}}

 \author[1,2]{Mattia Adamo}
 \author[3,2]{Michele Cantarella\thanks{Corresponding author. Email: \href{mailto:micant@dtu.dk}{micant@dtu.dk}}
}
\affil[1]{Institute of Cognitive Sciences Marc Jeannerod - CNRS}
\affil[2]{IMT School for Advanced Studies Lucca}
\affil[3]{Technical University of Denmark (DTU)}

\date{}

\maketitle

\begin{abstract}

\singlespacing{
Do workers behave the same in experiments as they do in real life? Combining a real-effort emoji-counting experiment with platform records, self-reported expectations and recalls, and smartphone screen-time logs, we compare how the same workers on Prolific respond to wage changes in the experiment and in their actual work. Our results reveal that workers' behavior in experiments is remarkably different from their behavior in real life. In the observational window, in fact, workers respond to expected wage changes just as they do in the experiment, but are much less responsive to unanticipated changes. However, once we condition on recall error in self-reports, we find that behavior in the two windows aligns. Recall error is then both a source of error in self-reported data and the result of a process of inattention that affects effort itself, as workers do not respond to unexpected wage changes that pass unnoticed.
}

\textbf{Keywords}: \emph{Labor supply elasticities, Intertemporal substitution, Expectations, Recall Error} 

\textbf{JEL codes}: J22, J31, D84, D90

\end{abstract}

\thispagestyle{empty}
\setcounter{page}{0}

\newpage

\section{Introduction}

Understanding how labor supply responds to changes in wages has long been a central concern in economics. The literature draws a fundamental distinction between labor supply responses to permanent and temporary wage changes, approximated by uncompensated and intertemporal elasticities, respectively. While the uncompensated (or Marshallian) elasticity reflects the impact of a permanent wage variation on labor supply affecting future income, the intertemporal (or Frischian) elasticity reveals how much workers are willing to alter the amount of time worked today when only today’s wage varies. These elasticities matter for a wide range of questions, from policy evaluation \citep{Keane2011} and incentive design \citep{Lazear2000} to micro-macro aggregation \citep{Chetty2011} and the analysis of market power \citep{Manning2020}, and are estimated in settings that differ widely in how wage changes arise and how workers perceive them.

The central theoretical difference between these elasticities lies in expectations. In life-cycle models \citep{pistaferri2003anticipated,Blundell1993,Altonji1988}, a wage change affects the intertemporal margin when workers perceive it as transitory relative to the wage path they expect in a generic working week, making work today more attractive than work in the future. It affects the uncompensated margin when it shifts their beliefs about future income, generating income effects that make work less attractive than leisure. In reference-dependent models \citep{Koszegi06,Camerer97,dunn1996loss,Goette2004}, expectations also act as reference points, so that the same wage change can have different effects depending on whether it is perceived as a gain or a loss. In applied research, however, these predictions often fail, as the distinction between these two elasticities blurs and estimates vary widely from study to study. These contradictions are especially evident between experimental and observational estimates, and ultimately reflect a well-known mismatch between micro and macro elasticities \citep{Chetty2011}. 

It is possible that responses to these expected and unexpected changes vary quite significantly between experimental and observational contexts. Typically, experiments implement wage changes that are explicitly temporary and highly salient, whereas in their actual work, workers may find temporary changes more difficult to track and only respond to persistent changes. Both life-cycle and reference-dependent models assume that workers keep track of all realizations, be they expected or unexpected, but evidence on limited attention suggests that this might not always be the case \citep{chetty2012bounds,gabaix2019behavioral,Bordalo2020}. Expectation levels and their deviations are also hard to measure, and most studies do not elicit them, but proxy them through the experimental design \citep[e.g.,][]{DOERRENBERG2023102305} or through other observables \citep{Thakral2021}. At the center of these inconsistencies, therefore, lie competing interpretations of the interaction between expectations and realizations that the empirical literature often ignores.

To our knowledge, no study has compared expectations and labor supply elasticities in \textit{both} observational and experimental contexts. In this paper, we address precisely this gap. Do workers behave the same in experiments as they do in real life? And how can we explain differences between these two windows? To answer these questions, we combine experimental and observational data from the same workers, connecting uncompensated and intertemporal elasticities within a unitary framework. These workers are active on Prolific, an online labor platform where researchers post studies at a set pay and workers choose which ones to take. Workers can then freely adjust their effort in response to available offers, facing the textbook continuous labor-leisure decision problem (examples include open-air market vendors, \citealp{Andersen2025}, cab drivers, \citealp{Camerer2022}, fishermen, \citealp{hammarlund2018trip}, bike messengers, \citealp{Fehr07}, and stadium vendors, \citealp{oettinger1999empirical}). As these workers earn income by routinely participating in studies and experiments like ours, the behavior observed in our study is a local realization of the same labor supply process that unfolds at the observational weekly horizon.

Our observational window focuses on how workers respond to anticipated and unanticipated wage changes during a reference working week. These estimates are not causal in a strict sense, but as we observe the same worker characteristics that clients and the platform use to determine task eligibility and pay, and absorb workers' propensity to work with individual fixed effects common to both windows, selection into better or worse paying tasks is largely accounted for, and the main threat to identification becomes measurement error rather than omitted variables. We address it by complementing labor supply self-reports with task completion logs from the platform registry and social media screen time statistics collected from respondents' devices, which do not share the same sources of recall error as self-reported measures. 

Our experimental window provides causal estimates of responses to anticipated and unanticipated wage changes in an experiment. The same workers are invited to take part in an emoji-counting real-effort experiment, where, over three sequences of tasks, they are asked to find and select specific emojis from a grid. The payout for the completion of a grid varies randomly between sequences, but its mean is anchored to workers' elicited expectations, which are randomized via an information treatment. With each successful completion, the grid gets larger while the payout stays the same, embedding diminishing returns to effort into the task, leaving workers free to decide when to skip a sequence and move to the next one. We exploit the variation in skipping decisions to estimate experimental labor supply elasticities.


As expectations are always collected, this setup allows us to estimate and compare uncompensated and intertemporal elasticities across both observational and experimental windows.
We find that workers' behavior in experiments is very different from that in their actual work. Intertemporal elasticities, in fact, are remarkably muted in the observational window when compared to the experiment, to the point that they are smaller than uncompensated elasticities, contradicting the predictions of life-cycle models. 

We find that this result is driven by a process of attention to realized earnings. Workers who keep track of their earnings and are able to recall them accurately respond to wage changes in the observational window just as they do in the experiment, while those who fail to track their earnings are much less responsive. Once we account for recall error, intertemporal elasticities also converge to the uncompensated ones, indicating that income effects are negligible. This finding suggests that recall error plays a dual role: first, as a source of error in self-reported data, and second, as a result of a process of inattention that affects effort itself when workers fail to track wage variations. It follows that expectations, and the attention workers pay to unexpected changes, are the primary drivers of behavior. Apparent cross-study heterogeneity in estimated elasticities may reflect differences in how salient deviations from expectations are, as unexpected wage changes often pass unnoticed.

\paragraph{Contribution to the Literature} Our results first contribute to the literature by helping reconcile the wide variation in wage elasticity estimates, especially between observational and experimental studies. While some of this variability reflects heterogeneity across workers \citep{BLUNDELL20074667}, and institutional \citep{Bargain2014} or historical \citep{Elder2025} contexts, a large part of it arises \textit{within workers}, from differences in how labor supply responses are defined and measured across studies, even when they try to capture the same type of elasticity. Sometimes, wage elasticities are studied across different time windows, over years \citep[see, for example,][]{Aaronson2009,Battisti2024,blomquist1983effect}, months \citep{Fehr07}, weeks \citep{Angrist2021,bourguignon1990labor}, days \citep{Farber2005,Camerer97}, or, in experimental settings, a few minutes \citep{Carpenter2016,Dickinson99,Keeley1978}. In other instances, when expected \citep{blundell1988labour} or desired \citep{cantarella2023task,Kahn1991, Altonji1988} hours of work are used in place of actual hours, it is the reference unit of effort that changes across studies. Estimates can also vary significantly depending on the mode of measurement, with self-reports and logged values often diverging because of measurement error \citep{Borjas1980,Barrett255}. We connect this source of variation with the literature on limited attention and optimization frictions. Agents underreact to incentives that are not salient \citep{chetty2009salience,gabaix2019behavioral}, and small optimization frictions can generate large differences between observed and structural labor supply elasticities \citep{chetty2012bounds}. Among gig workers, recalls of past pay are large and systematically biased, and closely linked to how workers form beliefs about their earnings \citep{Pires2025}. We show that once measurement error is addressed, uncompensated elasticities are similar across windows, and the remaining differences can be attributed to a recall channel, as workers who do not keep track of their earnings do not perceive deviations from their expected pay.

Our findings also corroborate the static gain-loss map by showing under which conditions labor supply is shaped by deviations from expectations. While classic personnel economics emphasizes that effort responds to pay schemes and incentive structures at the intensive margin  \citep[as documented in settings with performance pay, piece rates, and bonuses,][]{Lazear2000,Shearer2004}, the behavioral literature shows that such responses depend on reference points and expectations, which could greatly influence both short- and longer-term elasticities: mechanisms such as reference dependence and income targeting \citep{Camerer97, Abeler2011}, narrow bracketing \citep{MARTIN2017166}, and mental accounting \citep{Thaler1999} all incorporate expectations and have been invoked to explain some of the negative elasticities encountered in empirical research. Consistent with prospect theory \citep{kahneman2013prospect}, empirical evidence shows that identical wage changes can generate different effort responses depending on whether they are perceived as gains or losses \citep{DOERRENBERG2023102305,Goette2004,dunn1996loss}, and that deviations from reference wages matter more than absolute pay levels \citep{ZUBRICKAS2023176}. Similar results emerge when examining responses to variations in pay relative to the (perceived) wage and income of other workers \citep{Bracha2015,ZUBRICKAS2023176}. Nonetheless, evidence of reference dependence remains mixed, as studies such as \citet{Cosaert2022} find no evidence of income targeting even in controlled laboratory settings. Our results reconcile this literature by showing that reference dependence emerges when deviations from expectations are visible. In the experiment, where pay is displayed throughout the task, relative losses have about twice the effect of relative gains, consistent with loss aversion. In the reference week, workers do not respond to relative losses any more strongly than to relative gains.

We also contribute to a tangential body of literature on uncertainty and expectation formation in labor supply. This literature suggests that expectations are malleable and adjust over the day and week \citep{Thakral2021,Farber2008,JIA2021102004}, and that wage uncertainty \citep{Parker2005} may have a stronger effect on supply than wage levels. Putting uncertainty and expectations together, empirical evidence suggests that labor supply responses are in fact closer to neoclassical predictions when wages are known \citep{Fisher2025}, when individuals put less value on earnings \citep{Exley2019}, and when individuals are not loss-averse \citep{Fehr07}. Focusing on income effects, \citet{Andersen2025} find that initial unexpected income shocks trigger reference-dependent responses, but that workers quickly adapt and increase their supply once these shocks persist. We contribute to this literature by explicitly eliciting expectations with several questions, and evaluating how they influence supply over different margins and windows. We also show that, once stated, expectations remain the reference point throughout the experiment, as workers do not update it to the pay received in earlier sequences.

Beyond individual behavior, our results have implications for the aggregation of labor supply responses to the macro level \citep{Chetty2011,Erosa2016,Kneip2019}. If within-worker elasticities are similar across environments, differences in macro elasticities may reflect not preference heterogeneity but differences in how closely workers track their earnings. Our findings therefore also inform the design and evaluation of incentives within firms and organizations \citep{Lazear2000}, where compensation schemes may operate by making pay changes more or less visible, and the assessment of market power in monopsonistic settings \citep{Manning2020}, where firms’ ability to set wages, fees, and informational environments influences labor supply through what workers perceive as well as through prices.

The remainder of the paper is structured as follows. Section \ref{s:data} describes the data sources and the design of both the survey and the experiment. Section \ref{s:method} details our empirical strategy in each window, covering our randomization and instrumental variable strategies, and how we compare the two windows within the same workers. Section \ref{s:results} presents the results for the observational and experimental windows, their comparison, and the role of recall accuracy. Section \ref{s:conclusions} concludes, discussing our results and their implications.

\section{Data}
\label{s:data}

We combine survey, experimental, and platform data. We recruited participants on Prolific, an online labor platform widely used in academic research \citep{peer2017beyond}.


Online platforms like Prolific provide an ideal environment in which to study labor supply, and have been used both to estimate labor supply from observational data \citep{BARBOS2022382,Camerer2022} and to conduct behavioral experiments \citep{DOERRENBERG2023102305,Dube18,peer2017beyond,Berinsky2012,Horton2011,Paolacci2010}. 
Platforms break down jobs into small, discrete tasks with output-based pay, and workers, as independent contractors, are free to decide when and how much to work, facing the continuous labor-leisure choice of the textbook model. Prolific is particularly suited to our study, as its workers earn income by taking part in studies like ours. The supply decisions workers make in our experiment and during the week are therefore directly comparable and easily interpretable, as the choices they make in the experiment are the same as those they make in every other study they take during the week.

\begin{figure}[!bth]
    \centering
    \includegraphics[width=0.75\textwidth]{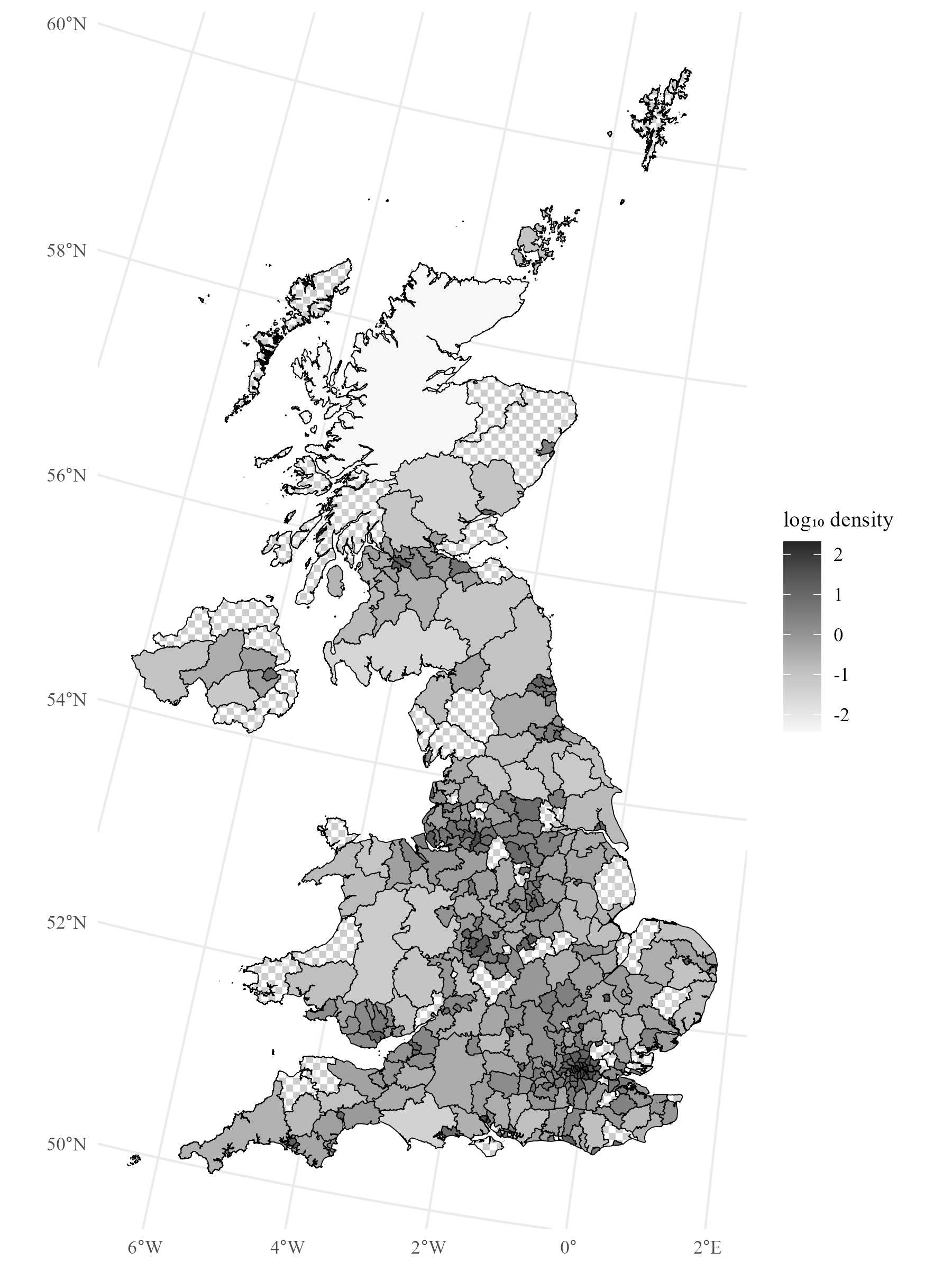} 
    \caption{Respondent Density by Local Authority District (UK)}
    \label{fig:Map_respondents}%
\caption*{\footnotesize{\textnormal{Notes: Logarithm of respondent density (respondents per 100 km$^2$) by Local Authority District, for the 1,632 study participants. Administrative boundaries correspond to the December 2021 ONS Local Authority Districts (UK BGC) dataset. 
Local Authority Districts with missing data are marked with a checkered pattern.}}}
\end{figure}

We recruited 1,632 study participants located in Great Britain.\footnote{Overall, 1,835 Prolific workers took part in the study without withdrawing from it. Of these, 169 did not pass the screening, 16 were rejected, and 21 timed out. As per Prolific's privacy policies, Prolific data on screened-out participants are unavailable.} The recruited sample is not intended to be representative of the general population, but should be broadly representative of the population where online platform work is especially prevalent (particularly in the Greater London area and the South of England), as shown in Figure \ref{fig:Map_respondents}. The survey fieldwork was conducted in two sessions: an initial pilot session between the 9th and 19th of December 2024, and a main session between the 13th of January and the 4th of May 2025. These dates were chosen so as not to overlap with any bank or public holidays. Responses were usually collected on Mondays and Tuesdays between 11:00 and 13:00 (GMT), in batches of around 50 respondents each, with a few exceptions for respondents who encountered technical issues and had to retake either the survey or the real-effort task.

Responses were screened along two dimensions: (i) mobile device type and (ii) recent activity on Prolific and online platforms. The device requirement was introduced to allow for the collection of screen time usage statistics, as we later detail. Since only Apple iPhone and Samsung Android devices provide standardized weekly usage reports, only participants using these devices were allowed to proceed. As of January 2025, these two brands accounted for over 80 percent of the mobile device market share in the UK,\footnote{At the time of the study, Apple and Samsung held an estimated 47.67 and 33.12 percent share, respectively. Source: Statcounter \url{https://gs.statcounter.com/vendor-market-share/mobile/united-kingdom/\#monthly-201003-202501}; last accessed: 27/01/2025.} which suggests that the screener is unlikely to introduce substantial concerns regarding external validity. 
The second screener admitted only participants who had been active on online platforms within the past month. This restriction was introduced to strengthen the internal validity of our measures, particularly the accuracy of self-reported recall of platform activity, and to ensure alignment with screen time data, which are also limited to the last month in both Apple and Samsung devices. The week during which participants last worked for Prolific is treated as the \textit{reference week} for the rest of the study. 
Due to this focus on regularly active users, findings should be generalized with caution to less engaged platform workers. 

The study was divided into two modules: a survey module for a fixed reward, and a real-effort task for a bonus reward linked to performance. Of all study participants, 95.65 percent also completed the real-effort task, for a total of 1,561 complete responses. Additionally, all study participants were asked to report the earnings and number of tasks logged by Prolific for their reference week, which we use to validate their self-reports.\footnote{This question was initially part of the survey, but we quickly found out that Prolific participants cannot access their submission records while a study is active. We therefore re-contacted survey respondents and collected this information in a separate study, in exchange for a fixed-rate payment, and later made it a compulsory step of the real-effort task.} 
Altogether, 1,377 study participants completed the full study and also validated their responses.\footnote{Not all of them enter every estimate. Each model uses the workers with non-missing values of its variables, and the comparison between windows is restricted to workers observed in both, excluding those whose target pay in the experiment was too large or too low, as we detail below. The number of observations is reported in each table.}

\newcommand{\msd}[2]{\num{#1} (\num{#2})}

\begin{table}[htbp]
\centering
\caption{Descriptive Statistics across modules}
\label{tab:desc}
\begin{adjustbox}{width=0.6\textheight,center}

\begin{tabular}{l c c c}
\toprule
 & \textbf{Survey Only} & \textbf{+ Task} & \textbf{+ Validation} \\
\midrule

\multicolumn{4}{l}{\textit{Descriptive Survey Statistics}} \\
Screentime (1 week ago)   & \msd{13.86}{13.16} & \msd{13.94}{13.11} & \msd{13.92}{13.26} \\
Screentime (2 weeks ago)  & \msd{13.76}{13.46} & \msd{13.84}{13.38} & \msd{13.83}{13.55} \\
Screentime (3 weeks ago)  & \msd{13.44}{13.13} & \msd{13.50}{13.08} & \msd{13.39}{13.08} \\
Expected screentime       & \msd{8.43}{8.54}   & \msd{8.43}{8.39}   & \msd{8.45}{8.51}   \\
Hours in uninterrupted session (expected) & \msd{2.29}{2.04} & \msd{2.28}{2.03} & \msd{2.25}{2.00} \\
Earnings in uninterrupted session (expected) & \msd{16.99}{18.76} & \msd{16.88}{18.56} & \msd{16.50}{18.13} \\
Tasks in uninterrupted session (expected)  & \msd{7.35}{7.50} & \msd{7.33}{7.55} & \msd{7.21}{7.46} \\
Earnings in 5 minutes (expected) & \msd{1.15}{0.87} & \msd{1.13}{0.85} & \msd{1.12}{0.83} \\
Tasks last week (actual)    & \msd{13.89}{9.83} & \msd{13.89}{9.81} & \msd{14.10}{9.69} \\
Tasks last week (expected)  & \msd{18.92}{12.78} & \msd{18.87}{12.71} & \msd{18.99}{12.58} \\
Earnings last week (actual) & \msd{23.87}{23.19} & \msd{23.95}{23.31} & \msd{23.26}{21.04} \\
Earnings last week (expected) & \msd{30.60}{27.70} & \msd{30.60}{27.71} & \msd{30.43}{27.20} \\
Hours of work last week (actual) & 9.05 (13.93) & 9.11 (14.09) & 9.45 (14.58) \\
Hours of work last week (expected) & 10.32 (15.26) & 10.38 (15.38) & 10.62 (15.62) \\
Risk propensity (1–32) & \msd{12.02}{8.71} & \msd{12.06}{8.73} & \msd{12.01}{8.82} \\
Impatience (1–32)                  & \msd{12.05}{10.97} & \msd{12.06}{10.95} & \msd{12.25}{11.00} \\
\addlinespace

\multicolumn{4}{l}{\textit{Platform logs}} \\
Years of Activity   & 2.70 (2.57) & 2.72 (2.58) & 2.71 (2.58) \\
Total Tasks Approved & 794.52 (898.9) & 800.19 (902.64) & 833.63 (921.01) \\
Tasks last week (real)       & -- & -- & \msd{13.42}{9.41} \\
Earnings last week (real)    & -- & -- & \msd{22.01}{20.88} \\

\addlinespace

\multicolumn{4}{l}{\textit{Demographic Statistics}} \\
Share employed     & \num{78.98}\% (\num{40.74}) & \num{79.37}\% (\num{40.46}) & \num{79.01}\% (\num{40.72}) \\
Working hours        & \msd{30.46}{14.32} & \msd{30.67}{14.15} & \msd{30.62}{14.01} \\
Share married      & \num{59.99}\% (\num{48.99}) & \num{60.09}\% (\num{48.97}) & \num{60.71}\% (\num{48.84}) \\
Share with degree  & \num{62.19}\% (\num{48.49}) & \num{62.08}\% (\num{48.52}) & \num{62.38}\% (\num{48.44}) \\
Age                  & \msd{37.70}{12.20} & \msd{37.79}{12.11} & \msd{38.09}{12.21} \\
Share female       & \num{59.75}\% (\num{49.04}) & \num{59.78}\% (\num{49.03}) & \num{60.07}\% (\num{48.97}) \\ 
\addlinespace

\multicolumn{4}{l}{\textit{Real Effort Task Statistics}} \\
\multicolumn{4}{l}{\textit{Tasks Completed}} \\
Sequence 1 & -- & \msd{11.47}{6.73} & \msd{11.39}{6.71} \\
Sequence 2 & -- & \msd{7.59}{5.85} & \msd{7.66}{5.86} \\
Sequence 3 & -- & \msd{7.13}{6.29} & \msd{7.25}{6.38} \\
\multicolumn{4}{l}{\textit{Earnings in £}} \\
Sequence 1  & -- & \msd{1.07}{0.90} & \msd{1.05}{0.88} \\
Sequence 2  & -- & \msd{0.74}{0.77} & \msd{0.74}{0.75} \\
Sequence 3  & -- & \msd{0.72}{0.87} & \msd{0.72}{0.88} \\
\multicolumn{4}{l}{\textit{Time on Task in minutes}} \\
Sequence 1  & -- & \msd{12.69}{17.63} & \msd{12.53}{17.78} \\
Sequence 2  & -- & \msd{9.47}{54.57} & \msd{9.96}{57.95} \\
Sequence 3  & -- & \msd{7.21}{17.07} & \msd{7.52}{17.96} \\
\multicolumn{4}{l}{\textit{Correct Responses}} \\
Sequence 1 & -- & \msd{10.22}{5.63} & \msd{10.14}{5.61} \\
Sequence 2 & -- & \msd{6.81}{5.07} & \msd{6.88}{5.07} \\
Sequence 3 & -- & \msd{6.48}{5.57} & \msd{6.57}{5.62} \\ \hline

Observations & 1,632 &  1,561 & 1,377 \\

\bottomrule
\end{tabular}
\end{adjustbox}

 \begin{tabular}{l}
\multicolumn{1}{p{0.95\textwidth}}{\footnotesize \textit{Note}: SD in parentheses.}\\
\end{tabular}

\end{table}

Table \ref{tab:desc} presents descriptive statistics for the three groups, which collectively exhibit no statistical discrepancy, alleviating our concerns about sample selection. 

The final dataset, along with the full study design, including the survey transcript and a standalone HTML file that replicates the real-effort task, is available as Supplementary Material.

\subsection{Survey, validation and screen time logs}
\label{sub:survey}

The survey module collected information primarily related to between-task labor supply and demographics, along with additional data used to elicit reference points, which are later used in the real-effort task. 

\begin{figure}[!bth]
    \centering

    \begin{subfigure}{0.45\textwidth}
        \centering
        \includegraphics[width=\linewidth]{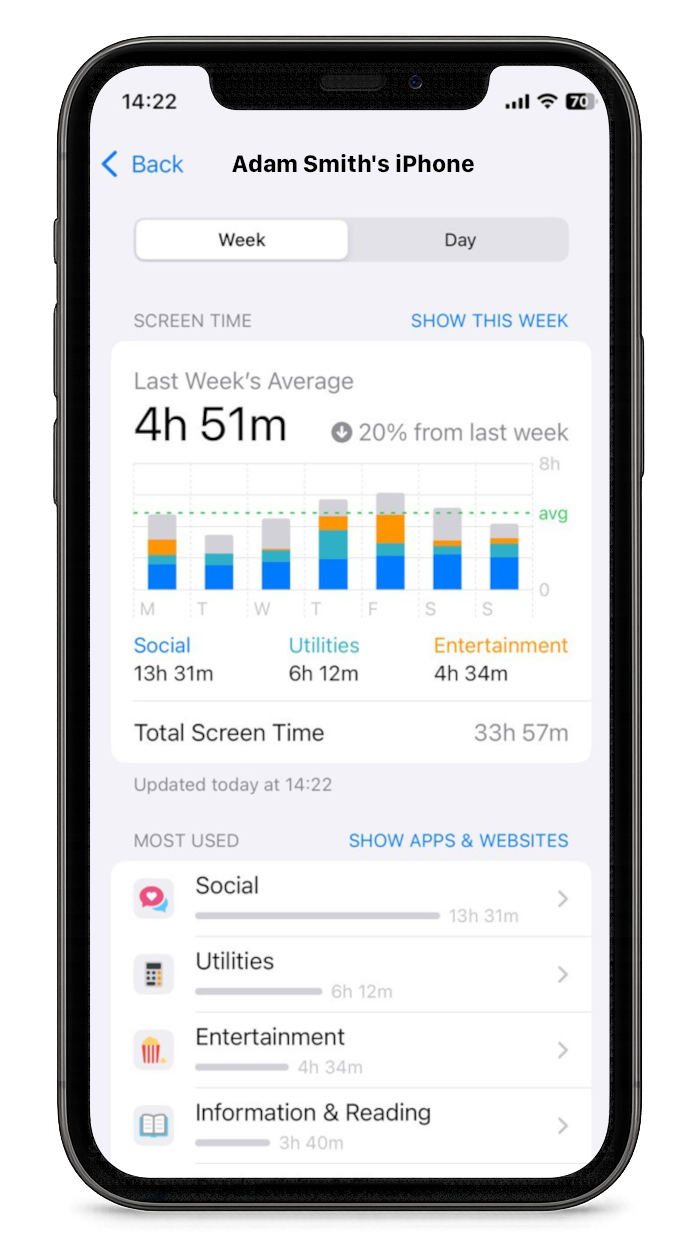}
        \caption*{\footnotesize{Apple iPhone}}
    \end{subfigure}
    \hfill
    \begin{subfigure}{0.45\textwidth}
        \centering
        \includegraphics[width=\linewidth]{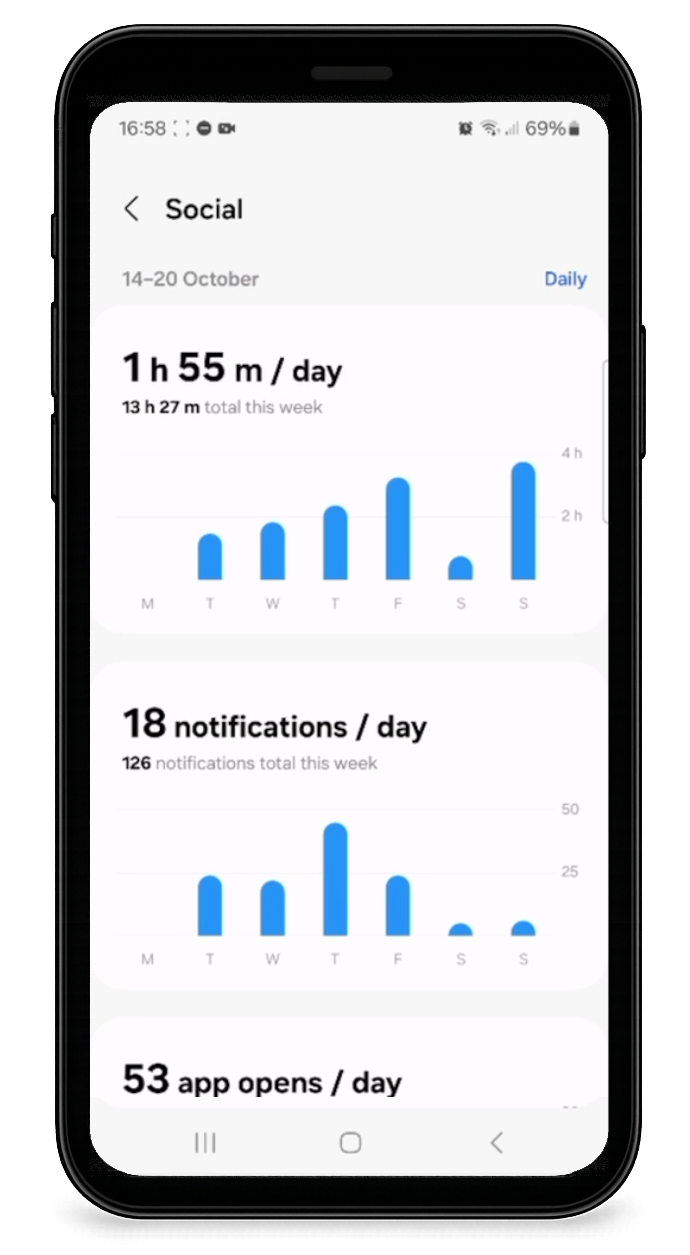}
        \caption*{\footnotesize{Samsung Android}}
    \end{subfigure}

    \caption{Weekly Social Media Screen Time}
    \label{fig:screen_time}
    \caption*{\footnotesize{\textnormal{Notes: Weekly social media screen time statistics on Apple iPhone and Samsung Android devices. Depending on their device choice, respondents were guided to either of these two screens via detailed instructions and a demonstrative video.}}}
\end{figure}

The module first captured screen time information. After completing the first two screeners, participants were asked to access and report the social media screen time logs of their primary smartphone for the past four weeks (including the current one) along with their stated screen time preferences. These measures serve as indicators of time use potentially correlated with the actual and expected work effort, while being largely free from self-reporting bias due to their basis in device-recorded data. We asked respondents to report their social media usage statistics rather than their total screen time, as this measure i) is the most preponderant leisure-related activity logged on a smartphone, and ii) is not confounded by platform and most other work-related activities, which might be captured by productivity and utility apps. Most importantly, while respondents may have heterogeneous preferences regarding the allocation of their free time between social media and other activities,\footnote{For example, some respondents might spend less time on social media relative to others, or might spend less time on social media in a given week relative to their habits (for example, they might spend the weekend traveling, or with their families and friends).} these preferences and activities are likely not correlated with the reporting error in their work hours, making these screen time statistics a reliable proxy for work hours.\footnote{We acknowledge that some workers might also be earning additional income from social media platforms (such as YouTube, Instagram, TikTok, or Facebook). Our estimates with the per-task wage, which rely on the number of tasks recorded by Prolific rather than on screen time, are very similar to those with the hourly wage, which suggests that this possibility is not a source of concern. Additionally, we reproduced our main results on the subsample of workers who reported income from Prolific as the sole source of platform income, and our main results were unchanged. These estimates are not included in the paper but are available upon request.} Screen time hours are then subtracted from the total of weekly hours (168) and linked with the corresponding reference week, collected through the second screener, to provide a measurement-error-free proxy of variation in observational working hours, as we later detail in Section \ref{s:method}. Figure \ref{fig:screen_time} provides a sample of the screen where respondents could find their social media screen time logs.


Following the collection of screen time information, we asked participants to indicate how regularly they were active on Prolific and other online platforms. We also asked which other labor platforms they were active on and which services they provided on these platforms. Additional information on workers' experience on Prolific was provided by Prolific's own pre-screeners, as reported below in Subsection \ref{sub:pdata}. The next part of the survey focused specifically on labor supply.  
This part of the survey was organized into several thematic blocks.

The first block captured information on reference points for an ideal work session. We began by eliciting unanchored reference points, as we deliberately provided no additional context and did not constrain participants’ answers to any specific range. We asked them to imagine an uninterrupted work session (with no breaks or waiting times) and to indicate how long such a session should last, how many tasks they would complete, and how much they would expect to earn for completing it. Next, participants were asked to consider an uninterrupted five-minute work session and report how much they would like to earn for it. This time, when eliciting these five-minute earnings, participants were randomly assigned to one of three anchors \citep[building from][]{KAHNEMAN1992175}, each displaying a different earning benchmark drawn from Prolific's own pay recommendations to researchers. The first group was shown the minimum hourly rate allowed (£6 per hour, equivalent to £0.50 for 5 minutes), the second group saw the average recommended rate (£9 and £0.75, respectively), and the third group received the maximum recommended rate (£12 and £1). These figures served as exogenous anchors in participants’ valuation of a short work session, allowing the researcher to address the potential endogeneity of the elicited reference points. As we detail in the next subsection, pay in the real-effort task was then randomized around these anchored expectations, unbeknownst to workers. 

The second block was designed to capture labor supply expectations and (recalled) realizations during the reference week.\footnote{We have been careful with the wording of these expectations, and asked participants who did not have a particular expectation in mind for the reference week to simply enter values they might find reasonable in a usual week.} We first recorded actual and expected earnings, tasks completed, and hours spent completing tasks and searching or waiting for tasks in the reference week. Self-reports from this block were then validated by the validation step: study participants who participated in this step were asked to provide exact logged information on their earnings and total number of completed tasks during their reference week by looking at their Prolific submission statistics.\footnote{Participants can access these statistics by accessing the link: \url{https://app.prolific.com/submissions}.} The difference between self-reported and validated earnings and tasks also provides our measure of how accurately workers recall their realized pay, which we use in Section \ref{s:results}. Finally, participants were asked the maximum temporary service fee, as a percentage of earnings, that they would tolerate before choosing not to work that week \citep{Mui2024,Kneip2019}. 

The rest of the survey captured important facts about our respondents' socio-demographics and economic preferences. The non-platform employment situation was thoroughly covered to capture all sources of income and potential labor supply constraints: participants reported the title and position of their main non-platform job (or their most recent one, if currently unemployed), and provided information on their monthly labor income and weekly working hours. Basic background information on respondents’ demographics and socio-economic status was also collected. The demographic section included standard items used in labor supply models, such as education, marital status, place of residence, household composition, and other employment-related variables. Data on age, gender, and nationality were instead provided directly by Prolific, as we later detail in Subsection \ref{sub:pdata}.

The next set of questions focused on non-labor income and household-level financial constraints. After capturing non-labor income sources, participants were asked how much they contributed to family expenses and how easily they made (or expected to make) ends meet at the time of the survey, one year earlier, and one year ahead. Then, patience and risk preferences were measured using a sequential staircase task, following the approach of \citet{Falk2018}.\footnote{Please refer to the online appendix of \citet{Falk2018} for the design of the staircase task.}

Upon completing the survey, respondents received a fixed payment of £2.25 and were then automatically redirected to the real-effort task.\footnote{At the end of the survey, the following debriefing text was shown: "You will now be redirected towards a behavioural experiment in which you can complete three sequences of tasks for a bonus reward. You will be able to skip each of these sequences when you feel you have earned enough bonus pay during a sequence. Please remember that we will only be able to pay you upon reaching the final screen of the behavioural experiment."}  

\subsection{Real-effort experiment}

The real-effort experiment aimed to observe workers' responses to changes in wages at the intensive and extensive margins. The experiment was designed to require participants to complete a trivial routine task, which nonetheless required their near-constant attention. 
The experiment was articulated through three sequences, each comprising an indefinite number of tasks,\footnote{While each sequence seemed to have no end, it actually comprised 42 tasks, for a total of 2 hours and 33 minutes.} following a 2-task practice sequence at the start of the session.\footnote{For better clarity, sequences were named "Sections" in the experiment.} Each task consisted of a simple emoji-clicking task \citep[drawing from the popular zero-counting task used in][]{Abeler2011} requiring participants to find and click inside a grid of emojis on replicates of a specific randomly selected emoji within an allotted time. 

Each sequence offered the same bonus payout upon successful completion of a task. However, each task in a sequence took increasingly longer to complete. In fact, the (initially 3x3) grid grew larger with each completed task, alternating with each new task between adding one emoji to each row and one to each column. The allotted completion time equaled 1 second per emoji in the grid plus a 6-second grace period.\footnote{In our initial pilot tests, we experimented with different intervals (0.5, 1, and 1.5 seconds per task), finding that the 1-second interval best allowed respondents to correctly evaluate each emoji in a grid without making mistakes while also minimizing the spare idle time at the end of a task.} The task was automatically submitted and evaluated only at the end of the allotted time, and participants could not submit it for evaluation beforehand. Upon submission of a task, the sequence proceeded to the next task, regardless of whether the response was correct. To prevent respondents from filling in the whole grid as soon as the task started, each row in the grid was progressively revealed in an interval of seconds equal to the number of emojis in each row. 

\begin{figure}[!bth]
    \centering
    \includegraphics[width=1\textwidth]{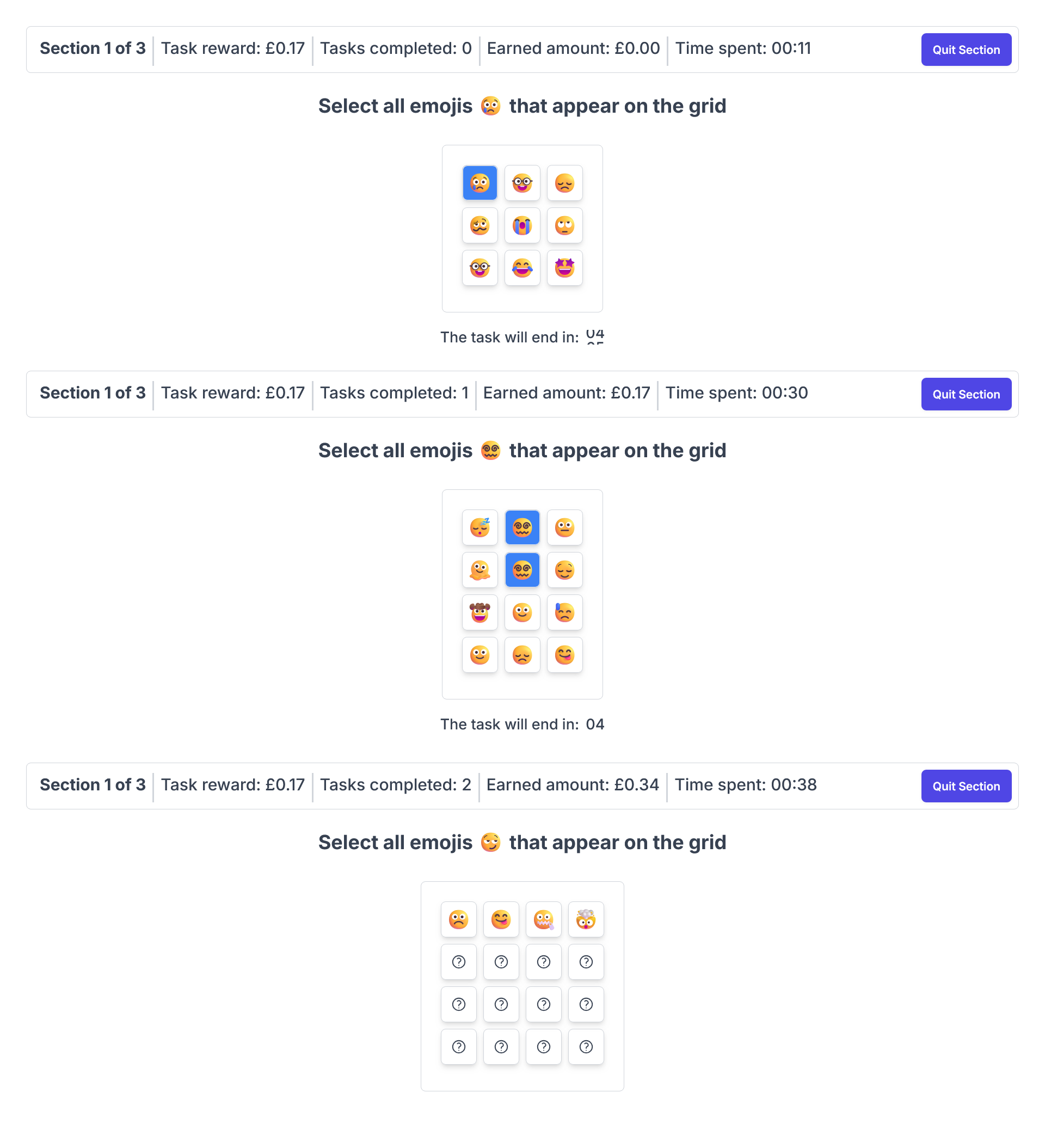} 
    \caption{The real-effort task}
    \label{fig:figure_experiment_hours}%
\caption*{\footnotesize{\textnormal{Notes: Sample emoji matrices used in the real-effort task for Tasks 1, 2, and 3, first sequence. Task 3 is shown mid-completion, illustrating the progressive row-revealing mechanism.}}}
\end{figure}

This setup implied that, within the same sequence, the marginal return to the time invested became progressively smaller with each task. Furthermore, while study participants were allowed to rest between sequences, each sequence could not be paused. Study participants could, however, terminate a sequence at any moment when they felt they had earned enough or spent enough time in it by clicking a "Quit section" button. By terminating a sequence, participants could cash in their earnings and move on to the next one, forfeiting all remaining uncompleted tasks (and associated rewards) from the current sequence.\footnote{Only 1 participant never pressed that button during a sequence, and also finished the entire study without skipping a single sequence, spending around 7 hours and 40 minutes in the study. When we contacted them to ask if they understood that they were expected to decide by themselves when to terminate a sequence, the participant said, "I completed all 3 sections because I enjoyed it."} This design relies on workers treating each sequence as an independent decision problem, substituting intertemporally within the sequence.

The payout associated with each sequence ranged from £0.01 to £0.22 per task. While these earnings were small enough that a single task could only have a negligible effect on workers' total income, income effects could still arise, as these earnings added to what workers earn on Prolific and were thus part of their income path. Payouts were randomized between sequences across three different levels, offered in random order: an endogenous \textit{target} pay based on idiosyncratic 5-minute earning expectations (we label this sequence as the \textit{target sequence}), and two exogenous levels set symmetrically above and below the target pay by a randomly assigned increment of 5, 10, 15, or 20 percent, allowing us to evaluate asymmetric responses to pay rate increases and decreases. The target pay was derived from respondents’ own answers in the survey module, allowing them to earn in five minutes approximately the same amount they reported expecting to earn in that time.\footnote{Since the allotted completion time for tasks was fixed, the target payout per task was computed by dividing the expected five-minute earnings by the task number $k$ at which total sequence completion time was closest to five minutes (this occurred at the end of task 9), and adjusting for the remainder.} As detailed in the previous section, random anchors were displayed when eliciting these expectations. We exclude from the experimental estimates the 290 respondents whose \textit{target} pay was either too large or too low.\footnote{The target pay was initially capped at £0.22 per task (around £2 in five minutes), and the cap was lowered to £0.12 (around £1.1 in five minutes, close to Prolific's maximum recommended rate) during the fieldwork. A total of 273 respondents reported five-minute expectations above the cap in place at the time and had their target pay capped, and a further 17, who reported expectations of £0.60 or less, had it set to £0.02.} 



A practice sequence was presented at the beginning of the study to familiarize participants with the real-effort task and the user interface. Concerning the latter, live statistics were also shown to study participants during each sequence, detailing the per-task unit reward and the current number of completed tasks, earnings, and time spent in the session, so that deviations from the target pay were always visible to workers. For each sequence, we recorded the same statistics, together with the number of correct tasks and the time spent idling, measured as the seconds in which respondents did not move their pointer, which we subtract from the time spent in the sequence to obtain active time.

Figure \ref{fig:figure_experiment_hours} shows sample screens corresponding to three tasks (task 1, 2, and 3) from a sequence, illustrating the mechanics and user interface outlined above. Other sample sections of the experiment are presented in Appendix \ref{A:experiment}. A standalone HTML file, which can be run locally to replicate the full real-effort task, is included as Supplementary Material.

\begin{figure}[!bth] 
\centering
    \includegraphics[width=\textwidth]{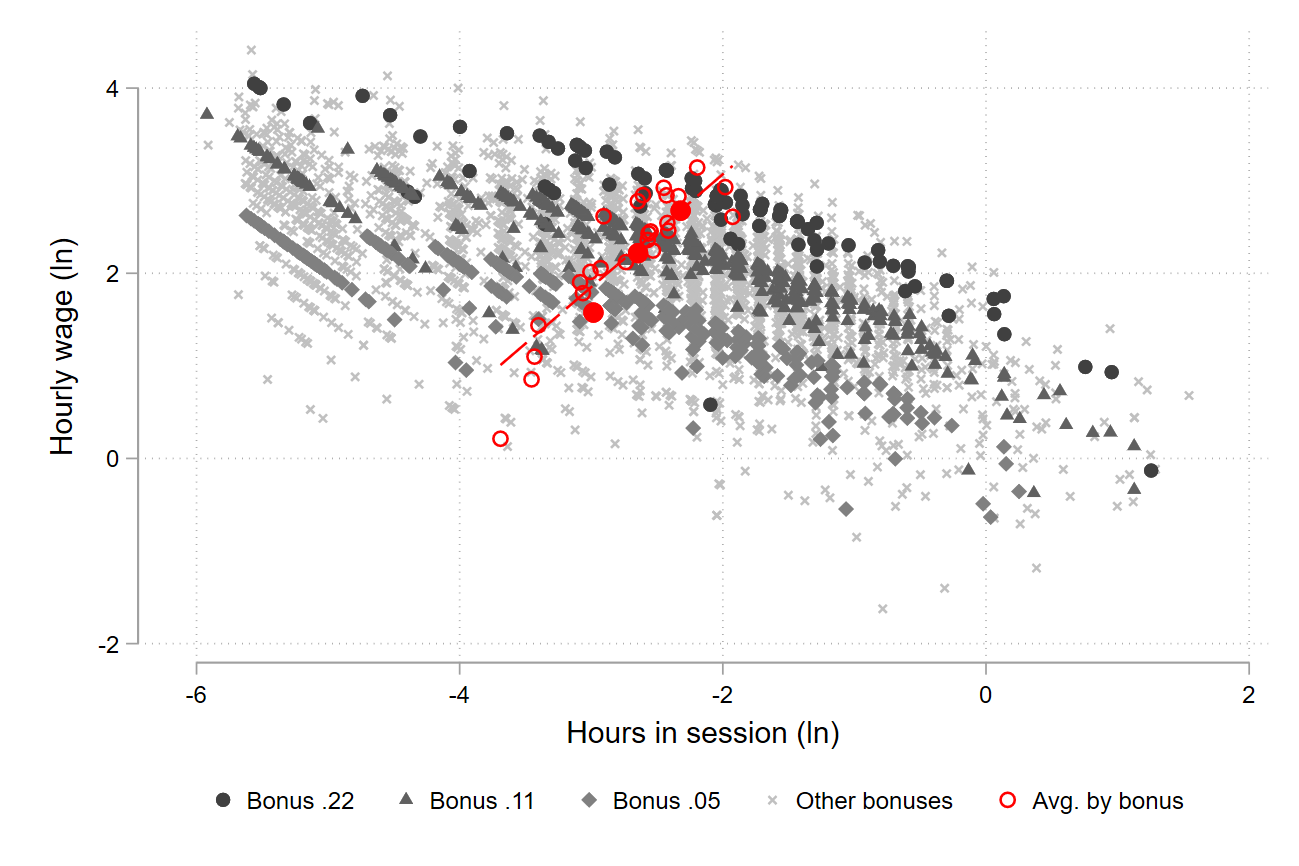}
    \caption{Labor supply curve in the experiment} \label{fig:experiment}
\caption*{\footnotesize{\textnormal{Notes: Each point is a sequence, with the total time spent in it and the resulting hourly wage, by pay per task. Within each pay level, the hourly wage falls with time, as each task takes longer than the previous one. Red circles are averages by pay level, and their linear fit (red line) traces the labor supply curve.}}}
\end{figure}

This particular setup fundamentally differs from the setup of other real-effort experiments \citep[such as][]{Dickinson99,Abeler2011} as we hard-code diminishing marginal returns from time spent working in the real-effort task. Previous research has instead relied on diminishing marginal utilities for the estimation of intensive margins, keeping the wage level fixed and measuring how long study participants kept working until they got bored. 
The diminishing returns to effort in the experimental task essentially force respondents into a different budget constraint for each sequence, each offering unique combinations of (potential) earnings and time investment due to the random variation in the payout, whose average level is anchored to the respondent's own expectations of 5-minute earnings. For each respondent, we observe three stopping points during three sequences. These represent three different utility maxima under three different budget constraints. Ideally, participants would terminate a sequence at the moment when they were indifferent between working and not working. By looking at how respondents' behavior at the intensive margin changes under different levels of payout, the econometrician can observe the labor supply curve of these workers and capture both intertemporal (between-sequence) and uncompensated (between-subject) margins. Figure \ref{fig:experiment}, showing time in session and hourly wage levels across sequences offering different payouts, provides a graphical intuition of this setup.

\FloatBarrier

\subsection{Prolific pre-screeners and registry data}
\label{sub:pdata}

Prolific allows researchers to access a wealth of supplementary information on study participants. Basic demographic information is always available, but researchers are also allowed to access all information that can be used to pre-screen study participants. 


Some of this information comes from the platform's registry. In fact, we were able to retrieve not only the lifetime number of approved Prolific submissions, but also the date of Prolific registration for each participant. These measures can help verify the self-reported labor supply measures collected through the survey. Furthermore, by dividing the logged total lifetime tasks, excluding those completed in the reference week, by the lifetime weeks of activity on Prolific, and weighting by regularity of use that we recorded in the survey, we obtained a measure of average tasks per week. %

The remaining Prolific data are also self-reported. Exogenous demographic information, such as age, gender, and place of birth, was also collected via the pre-screener set. Other endogenous pre-screeners, such as place of residence, marital status, education, and employment (among many others), were also included. 

Finally, through an agreement with Prolific, we obtained aggregate weekly statistics on the earnings and activity of all Prolific participants during the fieldwork period. This information includes, for each week, the average hours of activity, the average hourly wage, and the average number of completed tasks per week across all active Prolific participants. These figures allowed us to impute, for each reference week, a per-task actual wage by dividing real individual-level earnings by the average number of tasks per worker, which we use to instrument the per-task wage, as we detail in Section \ref{s:method}.

\section{Methodology}
\label{s:method}

Our empirical framework is designed to identify both the uncompensated (Marshallian) and the intertemporal (Frischian) labor supply elasticity for both experimental and observational windows. Any realized wage can be decomposed into the wage the worker expected and the deviation of the realized wage from this expectation. Variation in the expected wage is anticipated and shifts the expected income path of the worker, so that the response of effort to the expected wage is uncompensated, including both substitution and income effects. Deviations from the expected wage are unanticipated and, as long as workers expect their wage to return to its expected level in the following week, transitory, so that the response of effort to these deviations is intertemporal. Both margins can therefore be recovered from the same model, estimated on realized effort in each window.

\paragraph{Uncompensated and intertemporal elasticities} The general model is estimated as follows:
\begin{equation}
\label{eq:single}
    \ln(h_i) = \alpha + \Delta \ln(w_i) \, \eta_I + \ln(w_i^{E}) \, \eta_U + X_i^{'} \delta + \epsilon_i
\end{equation}

\noindent where $h$ is the \textit{effort} outcome, $w^E$ is the expected \textit{wage}, and $\Delta \ln(w_i) = \ln(w_i^{A}) - \ln(w_i^{E})$ is the deviation of the actual wage $w^A$ from its expected level. The parameter $\eta_I$ identifies the intertemporal elasticity, that is, how a one-percent unanticipated deviation from the expected wage translates into a percentage change in effort, holding expectations fixed. The parameter $\eta_U$ identifies the uncompensated elasticity, that is, how a one-percent increase in the anticipated wage translates into a percentage change in effort. Life-cycle models predict $\eta_I \geq \eta_U$, as income effects only follow anticipated wage changes \citep{MaCurdy1983,Keane2011}.\footnote{In the observational window, we also estimate first-difference specifications, which identify the intertemporal elasticity from the deviation of effort from expected effort. These estimates are discussed in Appendix \ref{A:levels}.}

Intertemporal elasticities can also be evaluated on the basis of the direction of the wage change relative to expectations:
\begin{equation}
\label{eq:sign}
    \ln(h_i) = \alpha + \Delta^{-} \ln(w_i) \, \eta^{-} + \Delta^{+} \ln(w_i) \, \eta^{+} + \ln(w_i^{E}) \, \eta_U + X_i^{'} \delta + \epsilon_i
\end{equation}

\noindent where $\Delta^{-} \ln(w_i)$ and $\Delta^{+} \ln(w_i)$ equal the deviation from expectations when this is, respectively, negative (a relative loss) or positive (a relative gain), and zero otherwise. Loss aversion with respect to expectations implies $\eta^{-} > \eta^{+}$.

\paragraph{Controls} Whenever not already absorbed by individual fixed effects, the vector of controls $X$ includes time-invariant individual controls predicting supply behavior and potentially correlated with the wage level. Among these, we include: (i) patience and risk propensity; (ii) contextual information on the time, day, and week of completion of the survey/experiment, proxying the worker's position along their current income path; and the standard set of controls entering a labor supply equation: (iii) non-platform income including unearned income and income from other sources, level of contribution to family expenses, and past and expected capacity to make ends meet; and (iv) a rich set of socio-demographic controls, including the Prolific pre-screeners that researchers use to set eligibility for their studies.\footnote{In our main specifications, these controls include age (and its squared term), gender, household size, marital status, years of experience on Prolific, years of experience in other occupations. Time, region, and last non-platform occupation and sector fixed effects are also included.}

\paragraph{Measurement and Estimation} Earnings and tasks in the reference week are measured with the \textit{real} values logged by the platform and collected in the validation step. As the platform does not log working time on an individual level, hours of work are \textit{self-reported}. In the experiment, effort is also measured as \textit{active} time, net of the time spent idling. Wages are measured on an \textit{hourly} or \textit{per-task} basis. \textit{Self-reported} earnings and tasks are also used to measure how accurately workers recall their realized pay, as we detail below. Table \ref{Table_dec_map} in Appendix \ref{A:fullest} summarizes the effort, wage, and instrument used in each specification.

All models are estimated with Ordinary Least Squares (OLS) and Two-Stage Least Squares (2SLS), as detailed in the next subsection.

\subsection{Observational window}

Estimates in the observational window are obtained by analyzing supply responses at the intensive margin during the reference week, combining survey, platform, and screen time data.

\paragraph{Uncompensated and intertemporal elasticities} Both elasticities are obtained from the general model in Equation \ref{eq:single}, and from Equation \ref{eq:sign} when the deviation is split by its sign. In the observational window, the effort outcome $h_i$ is the number of hours worked on Prolific during the reference week. The expected wage $w^E_i$, which identifies the uncompensated elasticity $\eta_U$, is computed as the ratio between expected earnings and expected effort in the reference week, with effort measured in hours (hourly wage) or in completed tasks (per-task wage). The deviation $\Delta \ln(w_i)$, which identifies the intertemporal elasticity $\eta_I$, is the log difference between the actual and the expected wage, where the actual wage divides real earnings by self-reported hours (hourly wage) or by real tasks (per-task wage).

\paragraph{Estimation} Observational elasticities cannot be estimated with Ordinary Least Squares (OLS) because of division bias \citep{Borjas1980}. When the effort term contains measurement error and is used to construct the wage, the wage inherits an error that is mechanically and negatively correlated with the outcome. This error affects all wage measures with self-reported hours or tasks in the denominator. Per-task wages can also be affected by a second source of division bias, arising from heterogeneity in task duration, as longer tasks yield higher pay at a given hourly rate.

Following the \textit{proxy IV} procedure from \citet{Borjas1980}, which \citet{gillen2019experimenting} generalize as \textit{obviously related IV}, we instrument each wage term by dividing earnings by a measure of effort that does not share the same error.\footnote{The procedure has seen extensive use in the literature. Among many, we cite \citet{Altonji86}, and among recent examples, \citet{card2016bargaining} and \citet{Fisher2025}.} Hourly wages are instrumented by dividing earnings by the weekly hours not spent on social media ($168$ minus the logged social media screen time), using the screen time of the reference week for the actual wage, and the average screen time of the other weeks for the expected wage.\footnote{Using the stated expected screen time for the expected wage yields nearly identical estimates.} Per-task wages are instrumented by dividing earnings by an average number of tasks: for the actual wage, the average number of tasks completed per worker on Prolific in the reference week, from the platform's aggregate statistics; for the expected wage, the worker's average weekly number of tasks, computed from the lifetime number of tasks recorded in the platform registry outside the reference week and weighted by how regularly the worker is active on Prolific, as detailed in Subsection \ref{sub:pdata}. The deviation $\Delta \ln(w_i)$ is instrumented by the difference between the instruments of the actual and the expected wage. In Equation \ref{eq:sign}, $\Delta^{-} \ln(w_i)$ and $\Delta^{+} \ln(w_i)$ are instrumented by the instrumented deviation, split by its own sign.

This procedure treats stated expected earnings as error-free, but these are elicited ex post and might be distorted by realizations in the reference week. As an alternative specification, we take advantage of the many proxies of expectations captured in our survey, each elicited with a different question, by using their common component as the instrument (which we denote as the composite instrument) \citep[following][]{kloek1960simultaneous,bai2010instrumental}. The aim of this procedure is to remove measurement error from the expected wage, not the component of expectations related to the worker's propensity to work, which is common to all measures and is addressed below. For the per-task wage, the measures are the instrument described above, the per-task wage the worker expects to earn in an uninterrupted work session, the time the worker expects to spend searching for studies per task completed, the stated five-minute target wage, and the temporary service fee on earnings that the worker states they would tolerate. For the hourly wage, we use the hourly counterparts of the first three measures: the instrument described above, the hourly wage expected in an uninterrupted work session, and the expected search time per hour of paid work. As each measure is elicited with a different question, their measurement errors are independent, while all of them reflect the pay the worker expects, so that each measure is a valid instrument for the others. 

The exclusion restriction underlying these procedures requires each instrument to be correlated with the true, error-free level of effort, and would fail if either instrument were driven by the same unobserved factors generating the measurement error in the wage term. For the screen time instrument, this could happen if skill, opportunity costs, leisure preferences, attention, procrastination, or home production affected both screen time and self-reported hours through a common channel; we view this as unlikely, since screen time is passively recorded by the device, independently of how respondents perceive or report their own effort. For the task-count instrument, the restriction requires that the heterogeneity in task length in the reference week is either common to the whole period of activity of the worker, and thus captured by the average number of tasks, or unrelated to the reference-period wage; we view this as plausible, since the pay of a study is set by the researcher on the basis of its estimated completion time, rather than of the time taken by the individual worker, unless the worker selected unusually long or short studies in the reference week. For the common component, the core assumption is that the reporting errors of the different measures are independent of each other and of the error in self-reported effort. This is more defensible than for any single instrument, as a violation would require all measures, elicited with different questions and in different units, to share the same reporting error.

It should be noted that the proxy IV approach is simply an error-correcting procedure, and a causal interpretation of the resulting estimates rests on the additional assumption that wages are uncorrelated with unobserved preferences for work. We argue that this assumption generally holds in our setting. In labor supply models, unobserved preference heterogeneity is generally treated as a second-order concern relative to measurement error in the wage \citep{Borjas1980}, once selection into tasks is controlled for \citep{Mroz1987}. On Prolific, the pay of each task is set by clients, who also use the platform's pre-screeners to decide which workers are eligible for it, so that workers' characteristics affect their wage only through eligibility. As we observe the same pre-screeners and include them as controls, we account for this selection margin directly. Similarly, expected wages should reflect the pay of the tasks workers are used to receiving, and thus the same eligibility margin.\footnote{Any remaining component of expectations related to workers' propensity to work can be absorbed by individual fixed effects when we pool the observational and experimental windows for the same workers (Subsection \ref{s:framework}).} Under these conditions, the observational and experimental elasticities should be quantitatively comparable, despite the former lacking strict causal identification. 

\subsection{Experimental window}

In the experimental window, we reshape our sample so that each sequence of the real-effort task is treated as its own point in time.

\paragraph{Uncompensated and intertemporal elasticities} To study both elasticities, we adapt the general Equation \ref{eq:single} as follows:
\begin{equation}
\label{eq:exp}
    \ln(h_{is}) = \alpha_{s} + \Delta \ln(w_{is}) \, \eta_I + \ln(w_i^{E}) \, \eta_U + X_i^{'} \delta + \epsilon_{is}
\end{equation}

\noindent where we add sequence constants $\alpha_{s}$ capturing the order in which a sequence is presented, and $h_{is}$ is the time spent by worker $i$ in sequence $s$. The expected wage $w^E_i$, which identifies $\eta_U$, is the target pay, that is, the per-task pay granting the worker's expected five-minute earnings in five minutes. As expected effort is fixed at five minutes, $\eta_U$ also indicates whether workers adhere to their expectations: a worker who stops after five minutes when offered the target pay would do so regardless of the level of the target, and $\eta_U$ would be zero. The deviation $\Delta \ln(w_{is})$, which identifies $\eta_I$, is the log difference between the pay offered in the sequence and the target pay, which is zero in the target sequence and equal to a random increment of 5, 10, 15, or 20 percent, above or below the target, in the other two. We also estimate Equation \ref{eq:exp} with individual fixed effects $\theta_i$, which absorb the expected wage, so that only $\eta_I$ is identified. Finally, as workers may update their reference point over the course of the experiment, in Appendix \ref{A:dynamic} we replace the stated expectation with the pay offered in a previous sequence.

\paragraph{Estimation} The intertemporal elasticity is identified by randomized deviations in Equation \ref{eq:exp}, with or without individual fixed effects, so that $\eta_I$ is causally identified without the need for instruments. Just as in the observational case, the uncompensated elasticity, when identified by the target pay, is instead derived from the worker's own expectations, which may be both endogenous and measured with error. 

The experiment provides an exogenous source of variation in the stated expectation. We instrument the target pay with the random anchor (detailed in Subsection \ref{sub:survey}) shown when the five-minute expectation was elicited. As the anchor only has three levels, we also use a stronger, composite instrument: the difference between the stated five-minute wage and the wage expected for an uninterrupted work session, which was elicited before the anchor was shown. Both wages depend on the worker's propensity to work, which cancels out in the difference, while the effect of the anchor remains. The logic is the reverse of the one we follow in the observational window, where we retain the component that different measures of the expectation have in common to remove measurement error: here, we remove the common component, which reflects the propensity to work, and only retain the random anchor and the component specific to the five-minute format.\footnote{Formally, let the stated five-minute wage be $\ln w^{5m}_{i} = \gamma A_{i} + \pi P_{i} + \varepsilon^{5m}_{i}$, where $A_i$ is the random anchor, $P_i$ is the propensity to work, and $\varepsilon^{5m}_{i}$ is the component specific to the five-minute format, that is, how the worker expects to adapt to short work sessions. The uninterrupted wage, elicited before the anchor, is $\ln w^{U}_{i} = \pi P_{i} + \varepsilon^{U}_{i}$. The instrument is $\ln w^{5m}_{i} - \ln w^{U}_{i} = \gamma A_{i} + \varepsilon^{5m}_{i} - \varepsilon^{U}_{i}$. Its correlation with the target pay comes from $\gamma A_i$ and $\varepsilon^{5m}_{i}$, which both enter the target pay and strengthen the first stage, while $P_i$ cancels out and $\varepsilon^{U}_{i}$, which is unrelated to the target pay, only adds noise to the instrument. Validity requires two assumptions that are stricter than those underlying the anchor alone, but reasonable in our setting. First, the propensity to work must enter both wages with the same weight $\pi$, so that it is fully removed by the difference. Second, $\varepsilon^{5m}_{i}$ and $\varepsilon^{U}_{i}$ must be uncorrelated with the error term of the effort equation, that is, the way a worker adapts to five-minute sessions affects time spent in the experiment only through the target pay. Both expectations are stated as earnings for a given session before the experiment starts, and, once the propensity to work is accounted for, the format-specific components only reflect how the worker values a short or an uninterrupted session, which in the experiment is captured by the target pay itself. Since the two instruments rely on different assumptions, the similarity of the resulting estimates (0.85 and 0.79, Table \ref{Table_2_exp}, columns 2 and 3) is reassuring.} 


\subsection{Comparison between windows}
\label{s:framework}

As we observe labor supply decisions for the same workers in both windows, we can compare their elasticities and test whether they respond to expected wages and to deviations from them in the same way in the experiment and in their actual work. Most importantly, observing each worker in both windows allows us to include individual fixed effects and partial out the endogenous component of expectations and realizations that is common to both windows.

\paragraph{Uncompensated and intertemporal elasticities} To estimate the joint elasticities, we stack each worker's reference week ($t=0$) and experimental sequences ($t=1,2,3$), and estimate Equations \ref{eq:single} and \ref{eq:exp} in the same regression:
\begin{equation}
\label{eq:within}
    \ln(h_{it}) = \theta_i + \tau_t + \sum_{k \in \{O,X\}} D^{k}_{t} \left[ \Delta \ln(w_{it}) \, \eta_I^{k} + \ln(w_{ik}^{E}) \, \eta_U^{k} \right] + \epsilon_{it}
\end{equation}

\noindent where $D^{O}_{t}$ and $D^{X}_{t}$ indicate the observational ($t=0$) and experimental ($t>0$) window, $\tau_t$ are period fixed effects, and $\theta_i$ are individual fixed effects. Each window keeps the effort outcome, the expected wage, and the deviation defined above, so that $\eta^{O}$ and $\eta^{X}$ are the elasticities of Equations \ref{eq:single} and \ref{eq:exp}, estimated on the same workers and in the same regression. Without individual fixed effects, and with the controls of each window, the estimates coincide with those of the separate models on this sample. Individual fixed effects absorb any determinant of effort that is common to both windows, such as time-invariant preferences for work and reporting behavior, and the component of expectations shared by the two windows, as workers who expect higher pay on Prolific are likely to state both a higher weekly expected wage and a higher five-minute target. Uncompensated elasticities are then identified from the window-specific component of expectations, that is, from how the difference in effort between the reference week and the experiment varies with the difference between the expectations held in each window, while intertemporal elasticities are still identified from deviations within each window. 

\paragraph{Estimation} We estimate Equation \ref{eq:within} on the workers observed in both windows, which allows us to test the differences between windows directly. Without individual fixed effects, we also include time-invariant controls, otherwise absorbed by the individual fixed effects, which we interact with the window. Wage terms are instrumented as in each window, with the composite instruments (results with the single instruments are reported in Appendix \ref{A:single}). With individual fixed effects, the target pay is not instrumented, as it does not vary across sequences and its variation is largely absorbed by the fixed effects, so the resulting estimate should be read as a lower bound. Standard errors are clustered by worker and by occupation and sector.

The two windows are directly comparable when the observational wage is hourly. Per-task wages are also comparable, as in both windows they measure the pay for a standard unit of work. In the experiment, this unit is a grid of emojis, which follows the same progression in every sequence and for every worker, while in the reference week it is an average study, once the measurement error described above is addressed. The difference between the two units shifts all log wages of a window by the same constant, which is absorbed by the window-specific constants.

\paragraph{Mismatches and attention to realized pay} Elasticities may still differ between the two windows for several reasons. A fundamental difference between the two windows is that, in the experiment, deviations are fully observed by the worker, while in the reference week they may be only partially perceived.

In the life-cycle model, the uncompensated elasticity equals the intertemporal one net of income effects, $\eta_U = \eta_I (1 - \kappa)$, where $\kappa \in [0,1]$ measures the strength of income effects \citep{MaCurdy1983,Keane2011}. This holds when workers observe their deviations, as in the experiment, where the pay per task is displayed throughout each sequence. In the reference week, workers know their expected wage, but need to keep track of the earnings and tasks accumulated over the week to infer their realized wage. Workers who do not keep track of them may only perceive a share $m \in [0,1]$ of the deviation, and adjust their effort to the wage they perceive rather than to the realized one. Only the response to deviations is then attenuated, so that the observational intertemporal elasticity becomes $m \, \eta_I$, while the uncompensated elasticity is unaffected. When deviations are poorly perceived, the uncompensated elasticity can therefore exceed the intertemporal one, especially when income effects are small.


This discussion leads to a prediction: the observational intertemporal elasticity increases with how closely workers track their realized pay during the week, while the experimental one does not. To test this prediction, we proxy how closely each worker tracks their realized pay during the week with the accuracy with which they recall it at the time of the survey, measured with the log difference between self-reported and platform earnings (or tasks) in the reference week. Workers who follow their earnings during the week can recall them accurately, while workers who do not track them can only report an approximate figure. If tracking affects behavior, and not only reporting, workers who recall their pay less accurately should also respond less to deviations from their expectations. We use the absolute value of this difference, $M_i$, winsorized at the 99th percentile, as a measure of how inaccurately workers perceive their realized pay.\footnote{Winsorization removes a few extreme recall errors, which would otherwise weaken the first stages of the interaction terms. Results without winsorization are very similar, but are not reported in this manuscript.} We interact every wage term in Equation \ref{eq:within} with $M_i$, and instrument the interactions of the observational wage terms with the interactions of their instruments. The coefficients on the wage terms are then the elasticities of a worker with perfect recall ($M_i = 0$), and the coefficients on the interactions measure how elasticities change with the recall error. As the pay per task is displayed during the experiment, the interactions with the experimental wage terms should be zero, providing a placebo test. As before, the model is estimated without and with individual fixed effects.

The role of recall accuracy follows from a theoretical assumption, but other characteristics may also moderate the elasticities, or be correlated with recall accuracy. We therefore test whether recall accuracy is the only moderator by letting the data select the moderators with a lasso among all the worker characteristics we observe (Appendix \ref{A:lasso}).

\FloatBarrier

\section{Results}
\label{s:results}

\begin{figure}[!p]
\centering
    \includegraphics[width=\textwidth]{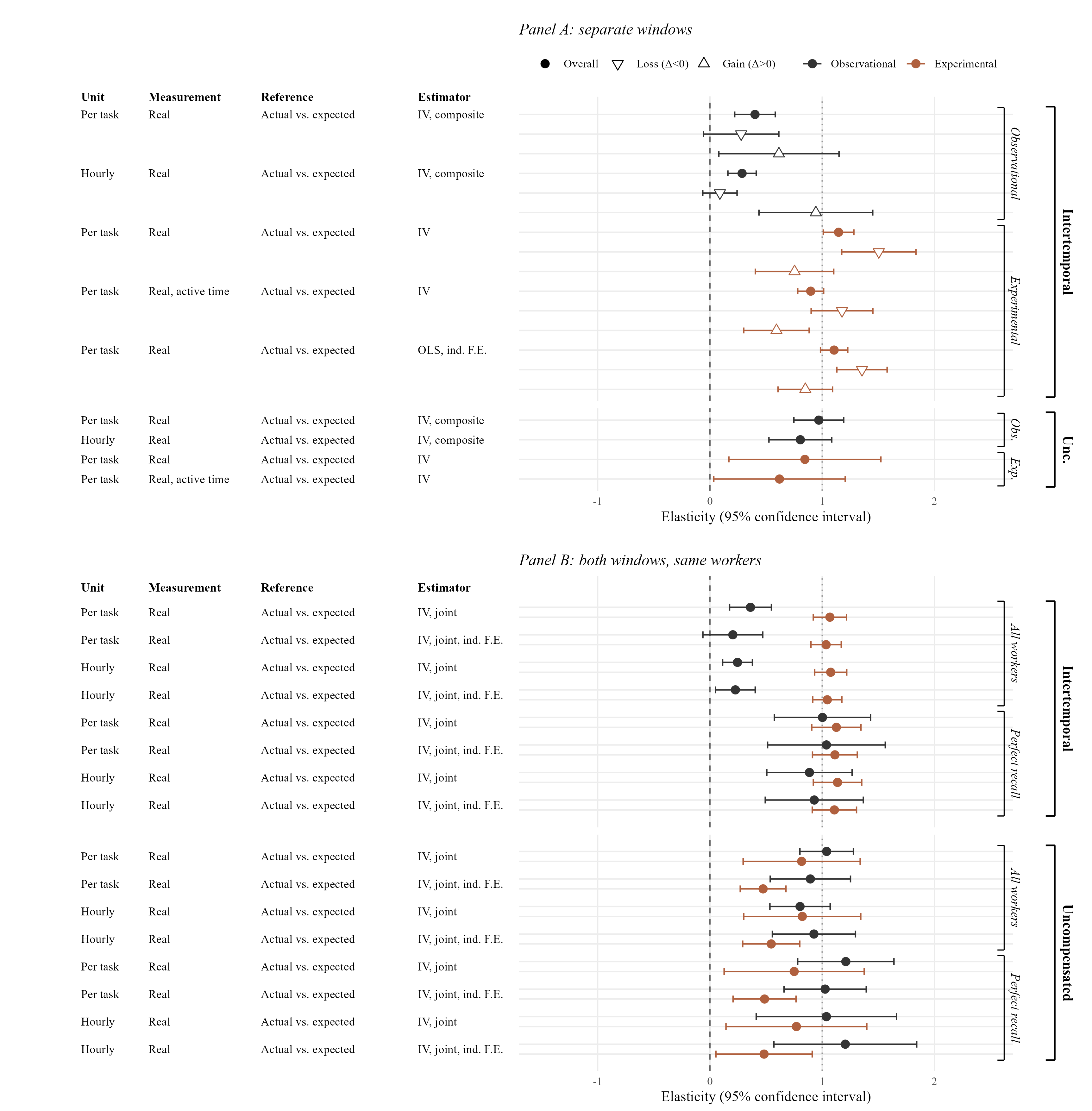}
    \caption{Labor supply elasticity estimates, observational and experimental windows}
    \label{fig:dec}
\caption*{\footnotesize{\textnormal{Notes: Point estimates and 95\% confidence intervals. Panel A: Tables \ref{Table_1_obs} (columns 3 and 6) and \ref{Table_2_exp}; losses and gains are intertemporal elasticities by sign of the deviation. Panel B: Tables \ref{Table_3_windows} (all workers) and \ref{Table_recall} (perfect recall of earnings, columns 1 to 4).}}}
\end{figure}

We present the results for the observational and experimental windows first separately, and then jointly. The forest plot in Figure \ref{fig:dec} summarizes our main estimates.

In the experiment, intertemporal elasticities are close to one (about 1.15) and larger than uncompensated elasticities (0.85), as predicted by life-cycle models. In the observational window, the ordering is reversed: uncompensated elasticities are 0.97 (per-task wage) and 0.81 (hourly wage), and intertemporal elasticities are 0.40 and 0.29. The observational intertemporal elasticity remains small in first-difference models (shown in Appendix \ref{A:levels}).

Later in this section, we compare the two windows within the same workers, allowing us to absorb the common component of expectations via individual fixed effects. We find that the uncompensated elasticity in the observational window is often statistically indistinguishable from the intertemporal elasticity in the experiment and from the experimental uncompensated elasticity, while the observational intertemporal elasticity is between 0.71 and 0.83 lower.

We then investigate the role of recall accuracy in explaining these residual differences. As predicted by our framework, differences in elasticities are driven by workers who do not accurately recall their realized earnings, as workers' responses to deviations from expectations depend on how accurately they recall their realizations.

\subsection{Observational window}

Table \ref{Table_1_obs} reports the estimates of Equations \ref{eq:single} (Panel A) and \ref{eq:sign} (Panel B), for the per-task wage (columns 1 to 3) and for the hourly wage (columns 4 to 6).

\begin{table}[!htbp] \centering 
  \caption{Labor supply elasticity estimates, observational window} 
  \label{Table_1_obs} 

\begin{adjustbox}{width=0.95\textwidth,center}

\def\sym#1{\ifmmode^{#1}\else\(^{#1}\)\fi}

\begin{tabular}{l*{6}{c}}
\hline\hline
                    &\multicolumn{1}{c}{(1)}&\multicolumn{1}{c}{(2)}&\multicolumn{1}{c}{(3)}&\multicolumn{1}{c}{(4)}&\multicolumn{1}{c}{(5)}&\multicolumn{1}{c}{(6)}\\
Effort:     &\multicolumn{6}{c}{Hours worked in the reference week (ln)}\\ \cmidrule(lr){2-7}
Wage:       &\multicolumn{3}{c}{Per task, real}&\multicolumn{3}{c}{Hourly, real}\\ \cmidrule(lr){2-4}\cmidrule(lr){5-7}
            & OLS & IV & IV & OLS & IV & IV \\
\hline
\multicolumn{7}{l}{\textit{Panel A: pooled}}\\
$\Delta$ Wage from expectations (ln): $\eta_I$&       0.140\sym{**}&       0.324\sym{***}&       0.401\sym{***}&      -0.512\sym{***}&       0.245\sym{**}&       0.286\sym{***}\\
                    &     (0.051)        &     (0.096)        &     (0.092)        &     (0.022)        &     (0.084)        &     (0.064)        \\
Expected wage (ln): $\eta_U$&       0.409\sym{***}&       1.194\sym{***}&       0.969\sym{***}&      -0.714\sym{***}&       1.421\sym{***}&       0.805\sym{***}\\
                    &     (0.059)        &     (0.114)        &     (0.113)        &     (0.023)        &     (0.214)        &     (0.143)        \\
p-value, $\eta_I = \eta_U$&       0.000        &       0.000        &       0.000        &       0.000        &       0.000        &       0.000        \\
\hline
\multicolumn{7}{l}{\textit{Panel B: $\Delta$ wage by sign of the deviation}}\\
Loss ($\Delta < 0$) &       0.118        &       0.288        &       0.278        &      -0.372\sym{***}&       0.240\sym{*} &       0.089        \\
                    &     (0.072)        &     (0.178)        &     (0.171)        &     (0.034)        &     (0.095)        &     (0.078)        \\
Gain ($\Delta > 0$) &       0.167        &       0.373        &       0.614\sym{*} &      -0.789\sym{***}&       0.254        &       0.943\sym{***}\\
                    &     (0.087)        &     (0.238)        &     (0.273)        &     (0.043)        &     (0.235)        &     (0.258)        \\
p-value, gain = loss&       0.689        &       0.817        &       0.406        &       0.000        &       0.960        &       0.004        \\
Observations with loss&         678        &         678        &         675        &         768        &         768        &         765        \\
Observations with gain&         645        &         645        &         643        &         556        &         556        &         552        \\
\hline
Instrument, actual wage:&          --        &Platform tasks        &Platform tasks        &          --        & Screen time        & Screen time        \\
Instrument, expected wage:&          --        &  Avg. tasks        &   Composite        &          --        & Screen time        &   Composite        \\ \hline
Time-Region F.E.    &         Yes        &         Yes        &         Yes        &         Yes        &         Yes        &         Yes        \\
Individual controls &         Yes        &         Yes        &         Yes        &         Yes        &         Yes        &         Yes        \\
SW F-test ($\Delta$ wage)&                    &     216.686        &     193.897        &                    &     248.226        &     442.290        \\
SW F-test (expected wage)&                    &     234.388        &     205.207        &                    &     122.386        &     132.950        \\
Adjusted R-Squared  &       0.206        &      -0.069        &       0.028        &       0.592        &      -2.755        &      -1.247        \\
Observations        &        1331        &        1331        &        1326        &        1327        &        1327        &        1320        \\
\hline\hline

\end{tabular}

\end{adjustbox}
 \begin{tabular}{l}
\multicolumn{1}{p{0.95\textwidth}}{\footnotesize Notes: SE clustered by occupation-sector in parentheses. Wages are computed with platform earnings and tasks; hours are self-reported. In the IV models, the effort denominators of the wage terms are replaced by the worker's average weekly tasks in the other weeks, from the platform registry (Avg. tasks), by the average tasks per worker on Prolific in the reference week (Platform tasks), or by the hours not spent on social media, from logged screen time (Screen time; for the expected wage, average of the other weeks). In columns 3 and 6, the expected wage is instrumented with a composite of earnings expectations elicited with different questions (Composite), their first principal component. In column 3 (6), these are the instrument of column 2 (5), the per-task (hourly) wage expected in an uninterrupted work session, the expected search time per task (per paid hour), the stated 5-minute target wage, and the tolerated service fee.}\\
\multicolumn{1}{l}{\footnotesize *p<.05, **p<.01, ***p<.001}\\
\end{tabular}

\end{table}

OLS estimates are affected by division bias, and are negative for the hourly wage ($-0.71$ for the uncompensated elasticity, column 4). All IV estimates are positive, and first stages are strong, with Sanderson-Windmeijer F-statistics above 104.7 \citep[the threshold suggested by][]{lee2022} for all instruments.

The uncompensated elasticity is 0.97 for the per-task wage and 0.81 for the hourly wage with the composite instrument (columns 3 and 6), and 1.19 and 1.42 with the single instruments (columns 2 and 5), all statistically significant at the 0.1\% level. The intertemporal elasticity is much smaller, at 0.40 and 0.29 with the composite instrument, and 0.32 and 0.25 with the single instruments. Different units of measurement and different instruments therefore yield similar estimates.

Panel B shows no statistically significant asymmetry between relative losses and gains, with elasticities of 0.28 and 0.61 for the per-task wage with the composite instrument (column 3), and of 0.29 and 0.37, and 0.24 and 0.25, with the single instruments (columns 2 and 5).\footnote{The exception is the hourly wage with the composite instrument, where the elasticity to relative gains (0.94) exceeds the one to relative losses (0.09), with a p-value of 0.004 for their difference (column 6). This asymmetry is opposite to the one found in the experiment, where losses have larger effects than gains.}

These results are at odds with the predictions of life-cycle models. In the observational window, the uncompensated elasticity is between two and six times larger than the intertemporal one (with p-values below 0.001 for their difference), while life-cycle models predict the opposite ordering, as income effects only follow anticipated wage changes. We return to this result when comparing the two windows below.

\FloatBarrier

\subsection{Experimental window}

Table \ref{Table_2_exp} reports the estimates of Equation \ref{eq:exp} (Panel A) and of its version by sign of the deviation (Panel B), for the time spent in the sequence (columns 1 to 3 and 7) and for active time (columns 4 to 6 and 8).

\begin{sidewaystable}[!htbp] \centering 
  \caption{Labor supply elasticity estimates, experimental window} 
  \label{Table_2_exp} 

\begin{adjustbox}{width=0.95\textwidth,center}

\def\sym#1{\ifmmode^{#1}\else\(^{#1}\)\fi}

\begin{tabular}{l*{8}{c}}
\hline\hline
                    &\multicolumn{1}{c}{(1)}&\multicolumn{1}{c}{(2)}&\multicolumn{1}{c}{(3)}&\multicolumn{1}{c}{(4)}&\multicolumn{1}{c}{(5)}&\multicolumn{1}{c}{(6)}&\multicolumn{1}{c}{(7)}&\multicolumn{1}{c}{(8)}\\
Effort:      &\multicolumn{3}{c}{Hours in the sequence (ln)}&\multicolumn{3}{c}{Active hours (ln)}&Hours (ln)&Active (ln)\\ \cmidrule(lr){2-4}\cmidrule(lr){5-7}\cmidrule(lr){8-8}\cmidrule(lr){9-9}
Expectation: &\multicolumn{6}{c}{Stated 5-minute target wage}&\multicolumn{2}{c}{Individual F.E.}\\ \cmidrule(lr){2-7}\cmidrule(lr){8-9}
             & OLS & IV & IV & OLS & IV & IV & OLS & OLS \\
\hline
\multicolumn{9}{l}{\textit{Panel A: pooled}}\\
$\Delta$ Wage from expectations (ln): $\eta_I$&       1.106\sym{***}&       1.146\sym{***}&       1.141\sym{***}&       0.872\sym{***}&       0.897\sym{***}&       0.904\sym{***}&       1.106\sym{***}&       0.884\sym{***}\\
                    &     (0.069)        &     (0.070)        &     (0.075)        &     (0.060)        &     (0.059)        &     (0.064)        &     (0.062)        &     (0.053)        \\
Expected wage (ln): $\eta_U$&       0.415\sym{***}&       0.846\sym{*} &       0.786\sym{***}&       0.346\sym{***}&       0.619\sym{*} &       0.699\sym{***}&                    &                    \\
                    &     (0.069)        &     (0.345)        &     (0.205)        &     (0.055)        &     (0.299)        &     (0.171)        &                    &                    \\
p-value, $\eta_I = \eta_U$&       0.000        &       0.376        &       0.069        &       0.000        &       0.348        &       0.216        &                    &                    \\
\hline
\multicolumn{9}{l}{\textit{Panel B: $\Delta$ wage by sign of the deviation}}\\
Loss ($\Delta < 0$) &       1.432\sym{***}&       1.504\sym{***}&       1.493\sym{***}&       1.130\sym{***}&       1.176\sym{***}&       1.186\sym{***}&       1.354\sym{***}&       1.045\sym{***}\\
                    &     (0.150)        &     (0.169)        &     (0.158)        &     (0.128)        &     (0.140)        &     (0.133)        &     (0.114)        &     (0.097)        \\
Gain ($\Delta > 0$) &       0.748\sym{***}&       0.754\sym{***}&       0.754\sym{***}&       0.589\sym{***}&       0.593\sym{***}&       0.596\sym{***}&       0.849\sym{***}&       0.718\sym{***}\\
                    &     (0.182)        &     (0.178)        &     (0.181)        &     (0.151)        &     (0.149)        &     (0.150)        &     (0.124)        &     (0.103)        \\
p-value, gain = loss&       0.024        &       0.018        &       0.015        &       0.032        &       0.027        &       0.019        &       0.013        &       0.053        \\
Observations with loss&        1221        &        1221        &        1219        &        1218        &        1218        &        1216        &        1485        &        1482        \\
Observations with gain&        1266        &        1266        &        1263        &        1264        &        1264        &        1261        &        1521        &        1519        \\
\hline
Instrument (expected wage):&          --        &      Anchor        &   Composite        &          --        &      Anchor        &   Composite        &          --        &          --        \\ \hline
Individual F.E.     &          No        &          No        &          No        &          No        &          No        &          No        &         Yes        &         Yes        \\
Sequence F.E.       &         Yes        &         Yes        &         Yes        &         Yes        &         Yes        &         Yes        &         Yes        &         Yes        \\
Time-Region F.E.    &         Yes        &         Yes        &         Yes        &         Yes        &         Yes        &         Yes        &          No        &          No        \\
Individual controls &         Yes        &         Yes        &         Yes        &         Yes        &         Yes        &         Yes        &          No        &          No        \\
SW F-test (expected wage)&                    &      74.192        &     130.350        &                    &      73.701        &     130.013        &                    &                    \\
Adjusted R-Squared  &       0.192        &       0.050        &       0.055        &       0.199        &       0.051        &       0.045        &       0.718        &       0.722        \\
Observations        &        3861        &        3861        &        3852        &        3855        &        3855        &        3846        &        4673        &        4666        \\
\hline\hline

\end{tabular}

\end{adjustbox}
 \begin{tabular}{l}
\multicolumn{1}{p{0.95\textwidth}}{\footnotesize Notes: SE clustered by respondent and occupation-sector in parentheses. Columns 1-6: the deviation is the randomized ratio between the bonus and the stated 5-minute target; the target is instrumented with the random anchor (columns 2 and 5) or with a composite instrument, the difference between the 5-minute and the uninterrupted-session wage (Composite, columns 3 and 6); workers whose target pay could not be set to their stated 5-minute expectation are excluded. Columns 7-8: the target is absorbed by individual fixed effects. Panel B: separate slopes for negative and positive deviations.}\\
\multicolumn{1}{l}{\footnotesize *p<.05, **p<.01, ***p<.001}\\
\end{tabular}

\end{sidewaystable}

The intertemporal elasticity is close to one in all specifications of time spent in the sequence, at 1.11 with OLS, 1.15 and 1.14 with the target pay instrumented with the random anchor and with the composite instrument (columns 1 to 3), and 1.11 with individual fixed effects (column 7). It is slightly smaller on active time, net of the time spent idling, between 0.87 and 0.90 (columns 4 to 6 and 8).

The uncompensated elasticity is smaller, at 0.85 and 0.79 with the two instruments (columns 2 and 3), about twice the OLS estimate of 0.41 (column 1), and at 0.62 and 0.70 on active time (columns 5 and 6). The anchor passes the conventional threshold for weak instruments (F-statistic of 74), but not the stricter one suggested by \citet{lee2022}, which the composite instrument passes (F-statistic of 130). As expected effort is fixed at five minutes, a positive uncompensated elasticity also indicates that workers overshoot their five-minute targets, and more so the higher their targets are.

Workers respond asymmetrically to the sign of the deviation (Panel B). Elasticities to relative losses are between 1.13 and 1.50 in columns 1 to 6, and those to relative gains between 0.59 and 0.75, with a difference statistically significant at the 5\% level in all six columns. With individual fixed effects, they are 1.35 and 1.05 for losses, and 0.85 and 0.72 for gains (columns 7 and 8).

These results are in line with life-cycle models, as the intertemporal elasticity exceeds the uncompensated one, in contrast with the observational window. With the instrumented target, the difference is not statistically significant (p-values of 0.38 and 0.07, columns 2 and 3), consistent with small income effects. When deviations from expectations are visible, workers respond to them elastically, and respond to relative losses roughly twice as strongly as to relative gains, consistent with loss aversion around the expected wage.

In Appendix \ref{A:dynamic}, we check whether workers update their reference point during the experiment, replacing the stated target with the pay of an earlier sequence. Our results suggest that they do not. When the deviation from the target and the change in pay from the previous sequence enter the same regression, only the deviation from the target affects effort (1.45, against 0.03), so that workers keep responding to their initial expectation throughout the experiment. Intertemporal elasticities are small in the first sequence (0.35), and rise to 1.49 in the following ones, increasing with the time spent in earlier sequences, consistent with workers learning how pay varies across sequences.

\FloatBarrier

\subsection{Comparison between windows}

Table \ref{Table_3_windows} compares the two windows within the same workers, following Equation \ref{eq:within}, without (columns 1 and 3) and with (columns 2 and 4) individual fixed effects.

\begin{table}[!htbp] \centering
  \caption{Labor supply elasticity estimates, observational and experimental windows within respondents}
  \label{Table_3_windows}
\def\sym#1{\ifmmode^{#1}\else\(^{#1}\)\fi}
\small
\begin{tabular}{lcccc}
\hline\hline
 & (1) & (2) & (3) & (4) \\
Effort: & \multicolumn{4}{c}{Hours of work (ln)} \\ \cmidrule(lr){2-5}
Observational wage: & \multicolumn{2}{c}{Per task, real} & \multicolumn{2}{c}{Hourly, real} \\ \cmidrule(lr){2-3}\cmidrule(lr){4-5}
Individual F.E.: & No & Yes & No & Yes \\
\hline
\multicolumn{5}{l}{\textit{Panel A: $\Delta$ Wage from expectations ($\eta_I$)}} \\
\quad Observational & 0.361\sym{***} & 0.204 & 0.246\sym{***} & 0.227\sym{*} \\
 & (0.095) & (0.136) & (0.068) & (0.090) \\
\quad Experimental & 1.068\sym{***} & 1.034\sym{***} & 1.076\sym{***} & 1.045\sym{***} \\
 & (0.075) & (0.069) & (0.073) & (0.066) \\
\quad p-value, observational = experimental & 0.000 & 0.000 & 0.000 & 0.000 \\
\addlinespace
\multicolumn{5}{l}{\textit{Panel B: Expected wage ($\eta_U$)}} \\
\quad Observational & 1.039\sym{***} & 0.894\sym{***} & 0.802\sym{***} & 0.926\sym{***} \\
 & (0.121) & (0.182) & (0.137) & (0.189) \\
\quad Experimental & 0.817\sym{**} & 0.473\sym{***} & 0.822\sym{**} & 0.546\sym{***} \\
 & (0.266) & (0.104) & (0.266) & (0.130) \\
\quad p-value, observational = experimental & 0.427 & 0.013 & 0.943 & 0.039 \\
\addlinespace
\multicolumn{5}{l}{\textit{Panel C: $\Delta$ Wage, losses ($\Delta < 0$)}} \\
\quad Observational & 0.177 & 0.095 & 0.025 & -0.014 \\
 & (0.186) & (0.294) & (0.096) & (0.129) \\
\quad Experimental & 1.416\sym{***} & 1.368\sym{***} & 1.398\sym{***} & 1.382\sym{***} \\
 & (0.165) & (0.149) & (0.161) & (0.153) \\
\quad p-value, observational = experimental & 0.000 & 0.000 & 0.000 & 0.000 \\
\addlinespace
\multicolumn{5}{l}{\textit{Panel D: $\Delta$ Wage, gains ($\Delta > 0$)}} \\
\quad Observational & 0.628\sym{*} & 0.383 & 0.832\sym{**} & 0.887\sym{**} \\
 & (0.268) & (0.354) & (0.267) & (0.301) \\
\quad Experimental & 0.697\sym{***} & 0.690\sym{***} & 0.709\sym{***} & 0.675\sym{***} \\
 & (0.189) & (0.152) & (0.184) & (0.152) \\
\quad p-value, observational = experimental & 0.820 & 0.450 & 0.690 & 0.548 \\
\addlinespace
\multicolumn{5}{l}{\textit{Panel E: Tests between margins}} \\
\quad p-value, observational $\eta_I = \eta_U$ & 0.000 & 0.000 & 0.000 & 0.000 \\
\quad p-value, experimental $\eta_I = \eta_U$ & 0.318 & 0.000 & 0.310 & 0.000 \\
\quad p-value, observational $\eta_U$ = experimental $\eta_I$ & 0.855 & 0.487 & 0.089 & 0.553 \\
\addlinespace Observations (respondents), Panels A-B, E & 4396 (1093) & 4416 (1097) & 4384 (1090) & 4404 (1094) \\
Observations (respondents), Panels C-D & 4372 (1098) & 4392 (1091) & 4372 (1098) & 4392 (1091) \\
\hline\hline
\end{tabular}
 \begin{tabular}{l}
\multicolumn{1}{p{0.95\textwidth}}{\footnotesize Notes: SE clustered by respondent and occupation-sector in parentheses. Observational wage terms instrumented as in Table \ref{Table_1_obs}, columns 3 and 6; target pay instrumented with the composite instrument of Table \ref{Table_2_exp} (columns 3 and 6) without individual F.E., and not instrumented with individual F.E. Estimates on respondents observed in both windows, excluding workers whose target pay was capped. Effort: hours worked in the reference week and time spent in each sequence. Without individual F.E.: the models of the two windows estimated jointly, each with its own controls and fixed effects, with all regressors interacted with the window. With individual F.E.: individual and period fixed effects, window-specific wage terms. Panels C-D exclude respondents with no deviation in the reference week.}\\
\multicolumn{1}{l}{\footnotesize *p<.05, **p<.01, ***p<.001}\\
\end{tabular}
\end{table}

Without individual fixed effects, uncompensated elasticities (Panel B) are 1.04 (per-task wage) and 0.80 (hourly wage) in the observational window, and 0.82 in the experimental window, so that the uncompensated elasticities of the two windows are statistically indistinguishable (the difference is 0.22 and $-0.02$). Intertemporal elasticities (Panel A) are 0.36 and 0.25 in the observational window, and 1.07 and 1.08 in the experimental window. The observational intertemporal elasticity is lower by 0.71 and 0.83, respectively, and both differences are statistically significant at the 0.1\% level. The uncompensated elasticity in the observational window and the intertemporal elasticity in the experiment are instead statistically indistinguishable (the difference is $-0.03$ and $-0.27$, with p-values of 0.86 and 0.09).

Adding individual fixed effects, the uncompensated elasticities become 0.89 and 0.93 in the observational window, and 0.47 and 0.55 in the experimental window, with differences of 0.42 and 0.38, statistically significant at the 5\% level. As the target pay is not instrumented in these models, the experimental uncompensated elasticity is attenuated by measurement error, and is best read as a lower bound. Intertemporal elasticities are 0.20 and 0.23 in the observational window, and 1.03 and 1.05 in the experiment, with differences of 0.83 and 0.82, both statistically significant at the 0.1\% level. The uncompensated elasticity in the observational window remains statistically indistinguishable from the intertemporal elasticity in the experiment (the difference is $-0.14$ and $-0.12$, with p-values of 0.49 and 0.55). Results are similar when the observational expected wage is instrumented with the single instrument of Table \ref{Table_1_obs} (columns 2 and 5), although the observational uncompensated elasticities are larger, between 1.00 and 1.52 (Table \ref{Table_3_windows_single}, Appendix \ref{A:single}). As individual fixed effects leave the estimates essentially unchanged, the differences between windows cannot be attributed to the composition of the sample, to time-invariant worker characteristics, or to the component of expectations shared by the two windows.

Panels C and D report elasticities by sign of the deviation. Without individual fixed effects, the experimental elasticity is 1.42 and 1.40 for relative losses and 0.70 and 0.71 for relative gains, while the observational elasticity is 0.18 and 0.03 for losses, not statistically significant, and 0.63 and 0.83 for gains. The observational elasticity is lower by 1.24 and 1.37 for losses, while for gains the difference is small ($-0.07$ and 0.12) and not statistically significant. With individual fixed effects, the experimental elasticity is 1.37 and 1.38 for relative losses, and 0.69 and 0.68 for relative gains, while the observational one is 0.10 and $-0.01$ for losses, and 0.38 and 0.89 for gains. The difference between windows is statistically significant at the 0.1\% level for losses, and not statistically significant for gains. The difference between windows is therefore concentrated in relative losses, to which workers respond strongly in the experiment and not at all in the reference week, while relative gains have similar effects in the two windows.
This comparison exercise highlights that the contradiction with life-cycle models persists in the observational window, even within the same workers. It also shows that the uncompensated elasticities are very close to, and often statistically indistinguishable from, the intertemporal elasticity in the experiment. This suggests, at a minimum, that income effects are small. In fact, if income effects were large, the uncompensated elasticity would be smaller than the intertemporal one in both windows. However, the gap between the intertemporal elasticities of the two windows remains large, so we still cannot draw conclusions about income effects from the observational window alone, where the intertemporal elasticity is smaller than the uncompensated one. The next subsection examines why.

\FloatBarrier

\subsection{Recall accuracy and labor supply}

If wage changes are less salient during the week than they are in the experiment, we would expect intertemporal elasticities to align once we control for the ability of workers to track these changes. Table \ref{Table_recall} tests this prediction, interacting all wage terms with the absolute recall error in earnings (columns 1 to 4) and in tasks (columns 5 to 8). We again begin with the models without individual fixed effects (odd columns).

For a worker with perfect recall of their earnings, the intertemporal elasticity in the observational window is 1.00 (per-task wage, column 1) and 0.89 (hourly wage, column 3), between two and four times the estimates of Table \ref{Table_3_windows}, and decreases by 0.39 and 0.32 for each log point of recall error, both statistically significant at the 0.1\% level. In the experiment, the intertemporal elasticity is 1.13 and 1.14, and is unaffected by the recall error, as expected given that pay is displayed during the task. For a worker with perfect recall, the difference between the observational and experimental intertemporal elasticity falls to $-0.12$ and $-0.25$, and is not statistically significant. The uncompensated elasticity in the observational window is 1.21 and 1.04 for a worker with perfect recall, and does not change significantly with the recall error, nor does the experimental one (0.75 and 0.77), from which it is statistically indistinguishable.

With individual fixed effects (columns 2 and 4), the observational intertemporal elasticity for a worker with perfect recall of their earnings is 1.04 and 0.93, and decreases by 0.54 and 0.38 for each log point of recall error, both statistically significant at the 0.1\% level. The experimental intertemporal elasticity is again unaffected by recall, and the difference between windows at zero recall error is $-0.08$ and $-0.18$, neither statistically significant. The observational uncompensated elasticity is 1.03 and 1.21 at zero recall error, and again does not change significantly with the recall error, while the experimental one (0.49 and 0.48) is unaffected by recall.

Columns 5 to 8 report the same models using the recall error in tasks. The observational intertemporal elasticity also decreases with the recall error (by between 0.24 and 0.45, statistically significant at the 5\% level), and the experimental interactions are again close to zero. Differences between windows for a worker with perfect recall of their tasks are larger (between 0.37 and 0.60), and statistically significant in three columns out of four, suggesting that what matters for perceiving deviations from expected pay is keeping track of one's earnings rather than of one's tasks. 

\begin{sidewaystable}[!htbp] \centering
  \caption{Labor supply elasticity estimates by recall accuracy, observational and experimental windows}
  \label{Table_recall}
\def\sym#1{\ifmmode^{#1}\else\(^{#1}\)\fi}
\begin{adjustbox}{width=0.95\textwidth,center}
\begin{tabular}{lcccccccc}
\hline\hline
 & (1) & (2) & (3) & (4) & (5) & (6) & (7) & (8) \\
Effort: & \multicolumn{8}{c}{Hours of work (ln)} \\ \cmidrule(lr){2-9}
Recall error: & \multicolumn{4}{c}{Earnings} & \multicolumn{4}{c}{Tasks} \\ \cmidrule(lr){2-5}\cmidrule(lr){6-9}
Observational wage: & \multicolumn{2}{c}{Per task, real} & \multicolumn{2}{c}{Hourly, real} & \multicolumn{2}{c}{Per task, real} & \multicolumn{2}{c}{Hourly, real} \\ \cmidrule(lr){2-3}\cmidrule(lr){4-5}\cmidrule(lr){6-7}\cmidrule(lr){8-9}
Individual F.E.: & No & Yes & No & Yes & No & Yes & No & Yes \\
\hline
\quad \textit{$\Delta$ Wage from expectations ($\eta_I$)} \\
\quad \quad Observational & 1.002\sym{***} & 1.037\sym{***} & 0.886\sym{***} & 0.929\sym{***} & 0.693\sym{***} & 0.530\sym{**} & 0.515\sym{***} & 0.487\sym{**} \\
 & (0.218) & (0.267) & (0.194) & (0.223) & (0.183) & (0.187) & (0.129) & (0.157) \\
\quad \quad \quad $\times$ recall error & -0.394\sym{***} & -0.537\sym{***} & -0.320\sym{***} & -0.379\sym{***} & -0.413\sym{*} & -0.451\sym{*} & -0.241\sym{**} & -0.247\sym{***} \\
 & (0.103) & (0.120) & (0.075) & (0.079) & (0.210) & (0.179) & (0.074) & (0.070) \\
\quad \quad Experimental & 1.126\sym{***} & 1.113\sym{***} & 1.136\sym{***} & 1.108\sym{***} & 1.080\sym{***} & 1.088\sym{***} & 1.085\sym{***} & 1.083\sym{***} \\
 & (0.112) & (0.102) & (0.110) & (0.101) & (0.119) & (0.109) & (0.118) & (0.108) \\
\quad \quad \quad $\times$ recall error & -0.079 & -0.104 & -0.082 & -0.090 & -0.015 & -0.083 & -0.012 & -0.055 \\
 & (0.105) & (0.096) & (0.104) & (0.099) & (0.130) & (0.121) & (0.130) & (0.125) \\
\quad \quad p-value, observational = experimental at zero recall error & 0.592 & 0.798 & 0.255 & 0.476 & 0.077 & 0.010 & 0.001 & 0.002 \\
\quad \textit{Expected wage ($\eta_U$)} \\
\quad \quad Observational & 1.210\sym{***} & 1.026\sym{***} & 1.037\sym{**} & 1.205\sym{***} & 0.930\sym{***} & 1.005\sym{***} & 0.746\sym{**} & 0.984\sym{***} \\
 & (0.219) & (0.187) & (0.319) & (0.324) & (0.267) & (0.220) & (0.233) & (0.259) \\
\quad \quad \quad $\times$ recall error & -0.156 & -0.146 & -0.140 & -0.218 & 0.114 & -0.199 & 0.051 & -0.108 \\
 & (0.146) & (0.149) & (0.176) & (0.172) & (0.284) & (0.266) & (0.172) & (0.191) \\
\quad \quad Experimental & 0.750\sym{*} & 0.486\sym{***} & 0.770\sym{*} & 0.482\sym{*} & 0.667 & 0.646\sym{***} & 0.691 & 0.512\sym{**} \\
 & (0.318) & (0.143) & (0.320) & (0.219) & (0.377) & (0.155) & (0.382) & (0.196) \\
\quad \quad \quad $\times$ recall error & 0.067 & -0.067 & 0.053 & 0.041 & 0.212 & -0.261 & 0.185 & 0.024 \\
 & (0.229) & (0.123) & (0.230) & (0.144) & (0.303) & (0.170) & (0.309) & (0.205) \\
\quad \quad p-value, observational = experimental at zero recall error & 0.273 & 0.016 & 0.570 & 0.020 & 0.549 & 0.135 & 0.900 & 0.074 \\
\addlinespace Observations (respondents) & 4388 (1102) & 4408 (1095) & 4376 (1099) & 4396 (1092) & 4396 (1104) & 4416 (1097) & 4384 (1101) & 4404 (1094) \\
\hline\hline
\end{tabular}
\end{adjustbox}
 \begin{tabular}{l}
\multicolumn{1}{p{0.95\textwidth}}{\footnotesize Notes: SE clustered by respondent and occupation-sector in parentheses. Estimates on respondents observed in both windows, as in Table \ref{Table_3_windows}, with the wage terms interacted with the absolute recall error. Recall error: log difference between self-reported and platform earnings (columns 1 to 4) or tasks (columns 5 to 8) in the reference week; absolute value, winsorized at the 99th percentile. Interactions of the endogenous terms are instrumented with the interactions of their instruments (the target pay as in Table \ref{Table_3_windows}), and elasticities are evaluated at a recall error of zero.}\\
\multicolumn{1}{l}{\footnotesize *p<.05, **p<.01, ***p<.001}\\
\end{tabular}
\end{sidewaystable}

Results are similar with the single instrument of the expected wage (Table \ref{Table_recall_single}, Appendix \ref{A:single}).
Taken together, these results support our prediction. In all specifications, the response to deviations from the expected wage in the reference week decreases with the recall error, while the response in the experiment does not, and for workers who accurately recall their earnings the difference between windows vanishes. The uncompensated elasticity, instead, does not depend on recall accuracy, as workers respond to the wage they expect whether or not they keep track of their realized pay. 

The same workers therefore have a single elasticity, close to one, which is revealed by their response to the anticipated wage in the reference week and by their response to deviations in the experiment, where deviations are visible. What differs between the two windows is the response to deviations from expectations, which is close to zero in the reference week, unless workers keep track of their earnings.

\paragraph{Robustness checks} In Appendix \ref{A:lasso}, we test whether characteristics other than recall accuracy moderate the elasticities, letting a lasso select the moderators among all the worker characteristics we observe. Out of 120 candidate interactions, the recall error in earnings is the only one selected as a moderator of the observational intertemporal elasticity, which it lowers by 0.46 (per-task wage) and 0.32 (hourly wage) for one standard deviation, while the experimental elasticities are unaffected. Among these characteristics, income targeting is the most obvious alternative, and for transparency we also report the estimates of Table \ref{Table_1_obs} with the wage terms interacted with indicators for workers who set weekly targets for earnings, hours of work, or the number of tasks (Table \ref{Table_targets}, Appendix \ref{A:fullest}). Setting a target does not change the response to deviations from expectations, contrary to what income targeting would imply.
\FloatBarrier

\section{Discussion and Conclusions}

\label{s:conclusions}

Expectations are a fundamental component of labor supply. 
They determine whether a wage change is anticipated or not, and thus which margin of labor supply it affects. 
In this paper, we estimated labor supply elasticities within the same workers, comparing uncompensated and intertemporal margins across both observational and experimental windows, and combining data from a real-effort experiment with survey self-reports, platform records, and screen time logs.


Our results suggest that, once measurement error is addressed, elasticities are unambiguously positive and close to unity at both margins. In the experiment, where deviations from expected pay are visible, the intertemporal elasticity exceeds the uncompensated one, as life-cycle models predict. In the reference week, the uncompensated elasticity is statistically indistinguishable from the intertemporal elasticity in the experiment, while the response to deviations from expectations is muted. These results hold even among workers who explicitly set earnings targets, and survive several robustness checks. 

What differs between the two windows is the attention workers pay to their realized pay. Workers who accurately recall their earnings respond to deviations from expectations in the reference week as strongly as in the experiment, while those who do not keep track of their earnings barely respond. As the observational intertemporal elasticity converges to the uncompensated one for workers who accurately recall their earnings, income effects also appear to be negligible. Recall error thus plays a dual role, as a source of error in self-reported data, and as a symptom of limited attention muting the response to deviations from expectations.


We also find clear evidence of reference dependence in labor supply, but only where deviations from expectations are visible. In the experiment, where pay is displayed throughout the task, relative losses with respect to the expected pay have about twice the effect of relative gains, consistent with loss aversion. In the reference week, losses and gains have instead similar effects.


%
How can these results be interpreted? Our results do not vindicate either of the standard neoclassical and reference-dependent interpretations of labor supply on their own. Workers appear to be neoclassical with respect to the wage they perceive. They respond to the wage they expect with an elasticity close to one, and income effects play a negligible role. Deviations from expectations, however, only matter when workers perceive them, and, when they do, relative losses weigh more than relative gains. When earnings are not tracked closely, most deviations from expected pay may simply go unnoticed, so that labor supply follows persistent changes in pay, which shape expectations, rather than transitory ones. In this sense, workers' behavior is neoclassical in its response to the wage they expect, and reference dependent in its response to deviations from it. When deviations are visible, losses weigh more than gains, but in their actual work, workers barely respond to relative losses.




Beyond these implications, these results have direct consequences for incentive design, operational decision-making, and the assessment of market power, especially in flexible labor markets, which are becoming increasingly common. A key implication of our findings is that workers' expectations, and how visible deviations from them are, are among the primary margins through which labor supply can be influenced. Firms, and especially platforms, that can shape expectations, or control how and when workers observe their earnings, effectively gain an additional source of leverage over labor supply and, in turn, market power, as temporary pay cuts that go unnoticed reduce labor supply less than they would if workers tracked their earnings. 

We conclude with a word about the limitations of our study and offer pathways for future research. The associations and causal effects that we establish are valid for a subpopulation of piece-rate online platform workers, and their validity should be extended to other contexts with caution, as it is possible that the income effect might become salient in other higher-wage contexts. Similarly, our observational window captures expectations anchored to a relatively short earnings cycle, and it is unclear whether the patterns we find hold over longer windows, where expectations are more persistent and income effects may play a larger role. Finally, and most importantly, we only show that recall accuracy moderates the response to deviations from expectations, not that attention causes it, as recall accuracy could be endogenous to unobserved worker characteristics affecting supply. Future research could establish this causal link by manipulating attention directly.

\bibliographystyle{myapalike}
\bibliography{references.bib}

\newpage
\appendix

\section*{Online Appendix}

\setcounter{table}{0}
\setcounter{figure}{0}
\renewcommand{\thetable}{A\arabic{table}}
\renewcommand{\thefigure}{A\arabic{figure}}
\renewcommand{\theHtable}{A\arabic{table}}
\renewcommand{\theHfigure}{A\arabic{figure}}

\section{Study instructions and practice sessions} \label{A:experiment}

In this section, we detail the instructions and prompts presented to study participants. Interested readers can also open the standalone HTML file with a web browser (available as Supplementary Material) to replicate the entire real-effort task locally.

\begin{figure}[!bth]
    \centering
    \includegraphics[width=0.65\textwidth]{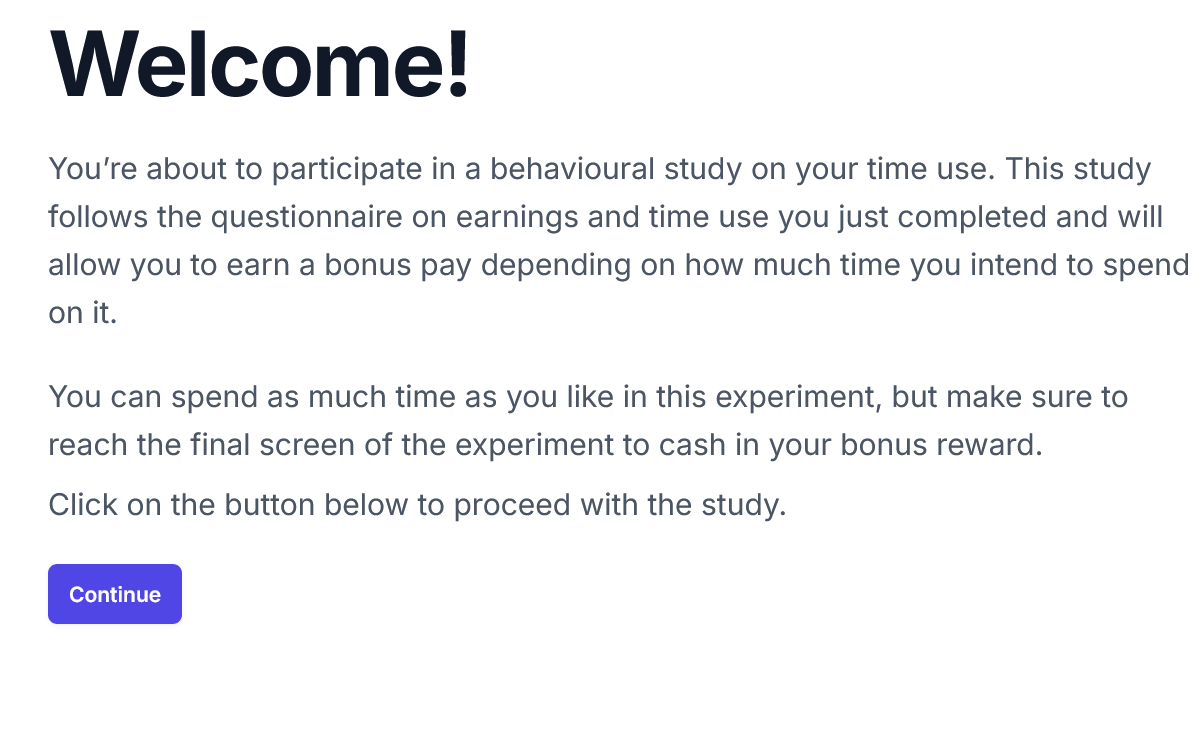} 
    \caption{Introduction screen}
    \label{fig:exp_introduction}%
\caption*{\footnotesize{\textnormal{Notes: Introduction screen presented to study participants upon joining the real-effort task.}}}
\end{figure}

\begin{figure}[!bth]
    \centering
    \includegraphics[width=1\textwidth]{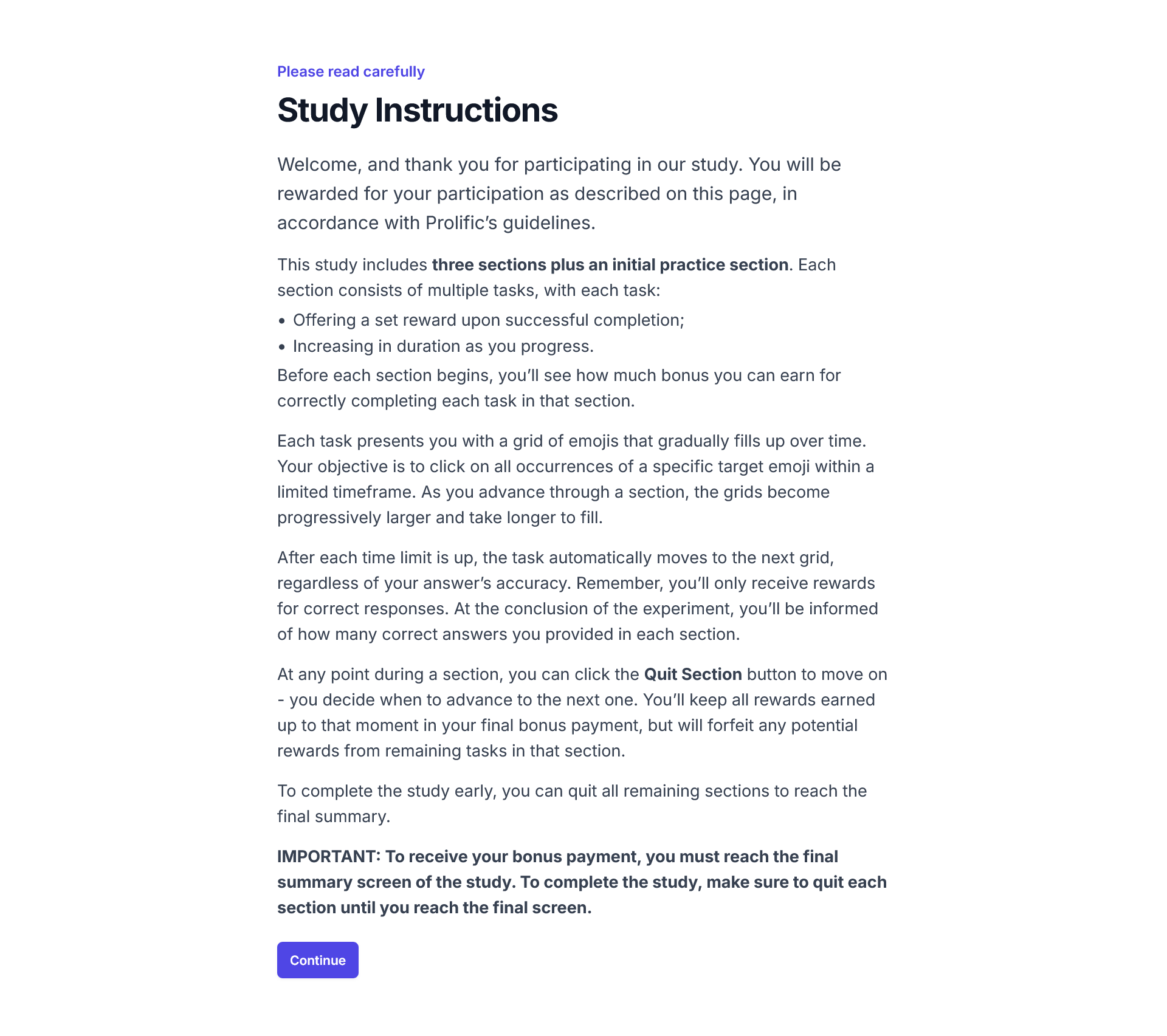} 
    \caption{Study Instructions}
    \label{fig:exp_instructions}%
\caption*{\footnotesize{\textnormal{Notes: Detailed study instructions presented to study participants.}}}
\end{figure}

Figure \ref{fig:exp_introduction} presents the introductory welcome screen shown to participants upon joining the study. Detailed instructions on the task are presented in the next screen, reported in Figure \ref{fig:exp_instructions}.

\begin{figure}[!bth]
    \centering
    \includegraphics[width=1\textwidth]{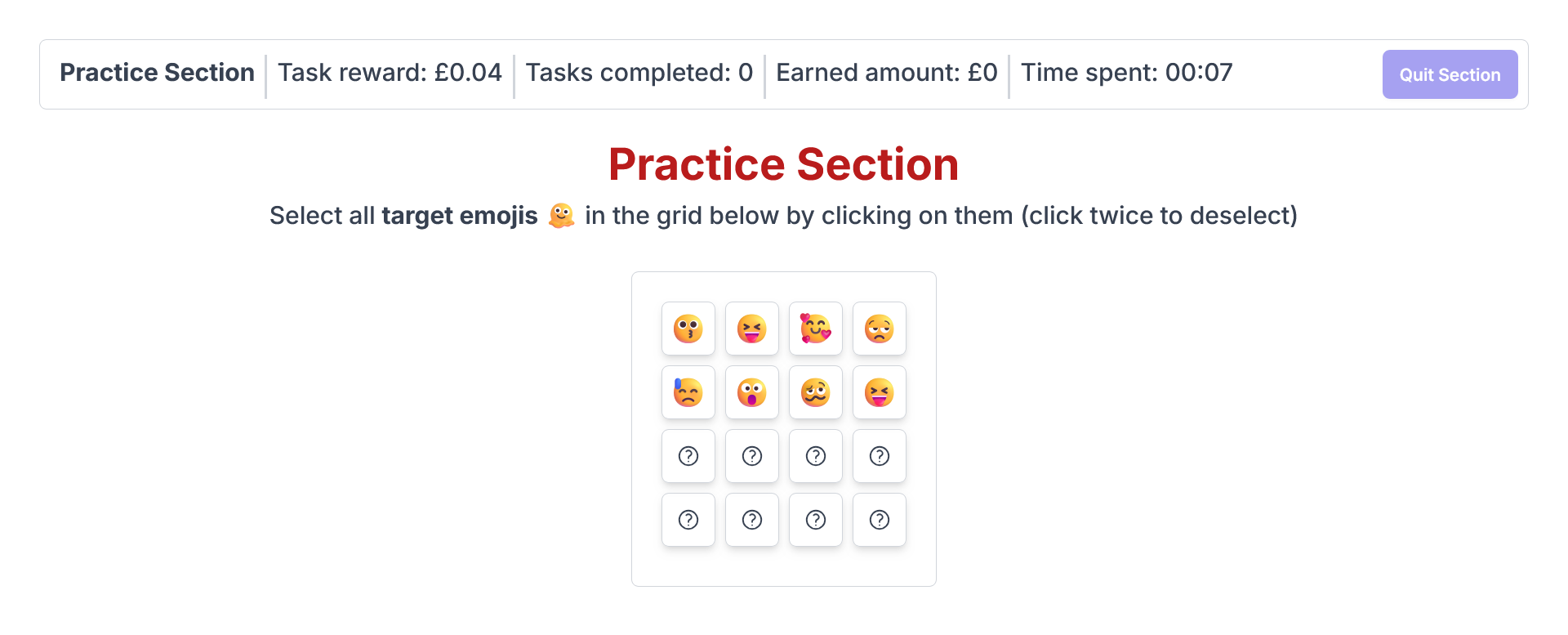} 
    \caption{1st Practice Task}
    \label{fig:exp_practice1}%
\caption*{\footnotesize{\textnormal{Notes: The first practice task consisted of a 4x4 grid, which filled over time. The task yielded a £0.04 payout upon successful completion, could not be skipped, and progress in the study was halted until the participant was able to select all correct emojis.}}}
\end{figure}

\begin{figure}[!bth]
    \centering
    \includegraphics[width=1\textwidth]{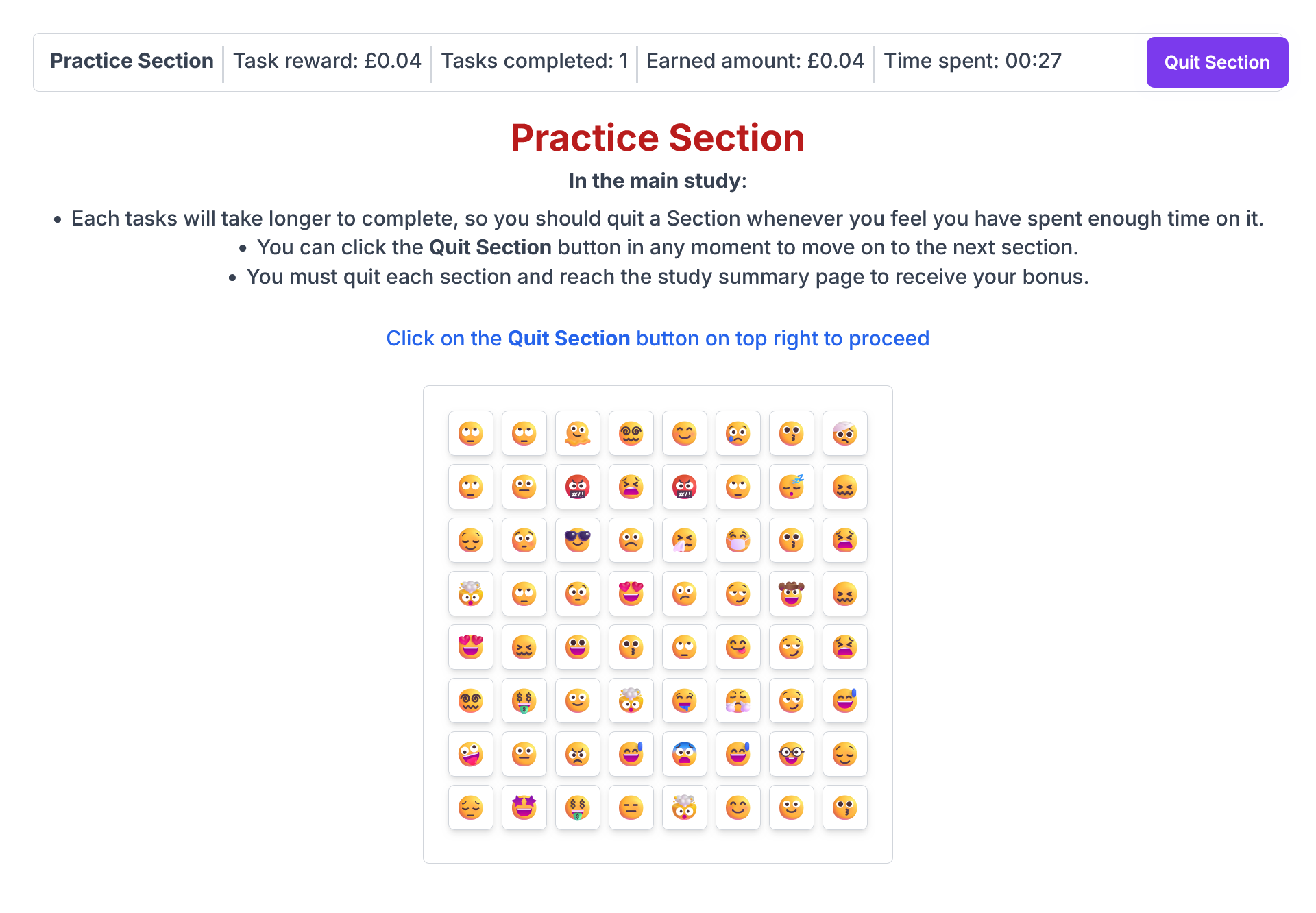} 
    \caption{2nd Practice Task}
    \label{fig:exp_practice2}%
\caption*{\footnotesize{\textnormal{Notes: The second practice task consisted of an already filled-up 8x8 grid. The task could not be completed (in fact, no prompt to click on a specific emoji was presented), and respondents were required to click on the "Quit Section" button to continue with the study.}}}
\end{figure}

\begin{figure}[!bth]
    \centering
    \includegraphics[width=0.75\textwidth]{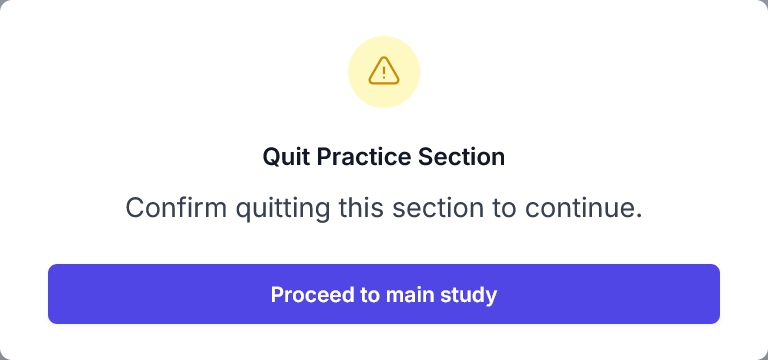} 
    \caption{Sequence quitting}
    \label{fig:exp_quit}%
\caption*{\footnotesize{\textnormal{Notes: Sequence quitting prompt presented to study participants upon clicking on "Quit Section".}}}
\end{figure}

Participants are then introduced to the practice sequence. Figures \ref{fig:exp_practice1} and \ref{fig:exp_practice2} show the two practice tasks, illustrating to study participants how to complete tasks and terminate work sequences. The prompt displayed upon quitting the second practice task is shown in Figure \ref{fig:exp_quit}. Once the practice session is complete, respondents move on to the three experimental sequences, as detailed in Section \ref{s:data} and shown in Figure \ref{fig:figure_experiment_hours} in the main text.

\FloatBarrier

\section{Additional results} \label{A:fullest}

\subsection{Specification map}

Table \ref{Table_dec_map} summarizes, for each column of Tables \ref{Table_1_obs} and \ref{Table_2_exp}, the effort outcome, the deviation and the expected wage, the estimator, and the instruments.

\begin{sidewaystable}[!htbp] \centering
\caption{Specifications in Tables \ref{Table_1_obs}, \ref{Table_2_exp}, and \ref{Table_2_exp_seq}}
\label{Table_dec_map}
\footnotesize
\begin{adjustbox}{max width=\linewidth}
\begin{tabular}{l p{3.6cm} p{5.2cm} p{3.6cm} c p{6.2cm} r}
\hline\hline
Col. & Effort (outcome) & $\Delta$ wage & Expected wage & Est. & Instruments & N \\
\hline
\ref{Table_1_obs}(1) & Hours, self-reported & Per-task, real (platform) / expected & Per-task, expected & OLS & Real earnings / platform average tasks in the week; expected earnings / average tasks & 1331 \\
\ref{Table_1_obs}(2) & Hours, self-reported & Per-task, real (platform) / expected & Per-task, expected & IV & Real earnings / platform average tasks in the week; expected earnings / average tasks & 1331 \\
\ref{Table_1_obs}(3) & Hours, self-reported & Per-task, real (platform) / expected & Per-task, expected & IV & Real earnings / platform average tasks in the week; composite of expected earnings / average tasks, session per-task wage, search time per task, 5-minute target, tolerated fee & 1326 \\
\ref{Table_1_obs}(4) & Hours, self-reported & Hourly, real earnings over self-reported hours / expected & Hourly, expected & OLS & Real earnings / non-screen time; expected earnings / non-screen time in other weeks & 1327 \\
\ref{Table_1_obs}(5) & Hours, self-reported & Hourly, real earnings over self-reported hours / expected & Hourly, expected & IV & Real earnings / non-screen time; expected earnings / non-screen time in other weeks & 1327 \\
\ref{Table_1_obs}(6) & Hours, self-reported & Hourly, real earnings over self-reported hours / expected & Hourly, expected & IV & Real earnings / non-screen time; composite of expected earnings / non-screen time in other weeks, session hourly wage, search time per paid hour, 5-minute target, tolerated fee & 1320 \\
\ref{Table_2_exp}(1) & Hours in the sequence & Bonus / 5-minute target (randomized) & 5-minute target wage & OLS & -- & 3861 \\
\ref{Table_2_exp}(2) & Hours in the sequence & Bonus / 5-minute target (randomized) & 5-minute target wage & IV & Randomized anchor & 3861 \\
\ref{Table_2_exp}(3) & Hours in the sequence & Bonus / 5-minute target (randomized) & 5-minute target wage & IV & Composite: 5-minute minus uninterrupted wage & 3852 \\
\ref{Table_2_exp}(4) & Active hours in the sequence & Bonus / 5-minute target (randomized) & 5-minute target wage & OLS & -- & 3855 \\
\ref{Table_2_exp}(5) & Active hours in the sequence & Bonus / 5-minute target (randomized) & 5-minute target wage & IV & Randomized anchor & 3855 \\
\ref{Table_2_exp}(6) & Active hours in the sequence & Bonus / 5-minute target (randomized) & 5-minute target wage & IV & Composite: 5-minute minus uninterrupted wage & 3846 \\
\ref{Table_2_exp}(7) & Hours in the sequence & Bonus / 5-minute target (randomized) & Absorbed by individual F.E. & OLS & -- & 4673 \\
\ref{Table_2_exp}(8) & Active hours in the sequence & Bonus / 5-minute target (randomized) & Absorbed by individual F.E. & OLS & -- & 4666 \\
\ref{Table_2_exp_seq}(1) & Hours in sequence 2 & Bonus 2 / bonus 1 (randomized) & Bonus in sequence 1 & OLS & -- & 1547 \\
\ref{Table_2_exp_seq}(2) & Hours in sequence 3 & Bonus 3 / bonus 2 (randomized) & Bonus in sequence 2 & OLS & -- & 1542 \\
\ref{Table_2_exp_seq}(3) & Hours in sequence 3 & Bonus 3 / bonus 1 (randomized) & Bonus in sequence 1 & OLS & -- & 1542 \\
\hline\hline
\end{tabular}
\end{adjustbox}
\end{sidewaystable}

\subsection{Weekly targets}

Income targeting models predict that workers who set a target for their weekly earnings stop working once they reach it, and therefore respond less, or even negatively, to deviations from their expected wage \citep{Camerer97}. In Table \ref{Table_targets}, we interact the wage terms of Table \ref{Table_1_obs} (columns 3 and 6) with indicators for workers who report setting weekly targets for earnings, hours of work, or the number of tasks. The intertemporal elasticity does not depend on any of the targets, with interactions between $-0.27$ and 0.24, none statistically significant. The uncompensated elasticity is also unaffected, with the exception of the hourly wage for workers with an earnings target ($-0.58$, statistically significant at the 5\% level). Workers with an earnings target work more hours at given wages, but do not respond differently to wage changes.

\begin{table}[!htbp] \centering 
  \caption{Labor supply elasticity estimates, observational window by target setting} 
  \label{Table_targets} 

\begin{adjustbox}{width=0.95\textwidth,center}

\def\sym#1{\ifmmode^{#1}\else\(^{#1}\)\fi}

\begin{tabular}{l*{6}{c}}
\hline\hline
                    &\multicolumn{1}{c}{(1)}&\multicolumn{1}{c}{(2)}&\multicolumn{1}{c}{(3)}&\multicolumn{1}{c}{(4)}&\multicolumn{1}{c}{(5)}&\multicolumn{1}{c}{(6)}\\
Wage:     & \multicolumn{3}{c}{Per task, real} & \multicolumn{3}{c}{Hourly, real} \\ \cmidrule(lr){2-4}\cmidrule(lr){5-7}
Target:   & Earnings & Hours & Tasks & Earnings & Hours & Tasks \\
\hline
$\Delta$ Wage from expectations (ln): $\eta_I$&       0.292\sym{*} &       0.429\sym{***}&       0.439\sym{***}&       0.243\sym{***}&       0.305\sym{***}&       0.296\sym{***}\\
                    &     (0.118)        &     (0.114)        &     (0.097)        &     (0.072)        &     (0.072)        &     (0.067)        \\
$\eta_I$ $\times$ target&       0.235        &      -0.109        &      -0.265        &       0.072        &      -0.093        &      -0.078        \\
                    &     (0.192)        &     (0.196)        &     (0.234)        &     (0.124)        &     (0.166)        &     (0.174)        \\
Expected wage (ln): $\eta_U$&       1.041\sym{***}&       0.957\sym{***}&       0.934\sym{***}&       0.990\sym{***}&       0.797\sym{***}&       0.786\sym{***}\\
                    &     (0.124)        &     (0.128)        &     (0.117)        &     (0.186)        &     (0.167)        &     (0.158)        \\
$\eta_U$ $\times$ target&      -0.298        &      -0.041        &       0.025        &      -0.577\sym{*} &       0.009        &       0.026        \\
                    &     (0.236)        &     (0.453)        &     (0.442)        &     (0.287)        &     (0.601)        &     (0.505)        \\
Target              &       0.425\sym{***}&       0.100        &       0.084        &       1.185\sym{**}&      -0.039        &       0.014        \\
                    &     (0.127)        &     (0.204)        &     (0.218)        &     (0.392)        &     (0.837)        &     (0.686)        \\
\hline
Time-Region F.E.    &         Yes        &         Yes        &         Yes        &         Yes        &         Yes        &         Yes        \\
Individual controls &         Yes        &         Yes        &         Yes        &         Yes        &         Yes        &         Yes        \\
SW F-test ($\Delta$ wage $\times$ target)&     193.061        &     162.380        &     143.697        &     429.462        &     182.048        &     112.146        \\
Observations        &        1326        &        1326        &        1326        &        1320        &        1320        &        1320        \\
\hline\hline

\end{tabular}

\end{adjustbox}
 \begin{tabular}{l}
\multicolumn{1}{p{0.95\textwidth}}{\footnotesize Notes: SE clustered by occupation-sector in parentheses. IV estimates of the specifications of Table \ref{Table_1_obs}, with the wage terms interacted with an indicator for respondents who report setting a weekly target for earnings, hours of work, or number of tasks. Interactions instrumented with the interactions of the instruments.}\\
\multicolumn{1}{l}{\footnotesize *p<.05, **p<.01, ***p<.001}\\
\end{tabular}

\end{table}

\subsection{First-difference models} \label{A:levels}

In the observational window, we also estimate first-difference specifications, which identify the intertemporal elasticity from the deviation of effort from its expected level. We regress the deviation of hours worked from expected hours on the deviation of the wage from the expected wage, $\ln(h_i) - \ln(h_i^E) = \alpha + \Delta \ln(w_i)\, \eta + \ln(w_i^E)\, \eta_E + X_i^{'} \delta + \epsilon_i$, with the same wage measures, instruments, controls, and fixed effects as Table \ref{Table_1_obs}. If expected hours respond to the expected wage with the same elasticity as actual hours, the expected wage drops out of the outcome ($\eta_E = 0$), and $\eta$ identifies the intertemporal elasticity.

Table \ref{Table_dod_obs} reports the estimates. The elasticity to the deviation is small in all IV specifications, between 0.04 and 0.11, even smaller than the observational intertemporal elasticities of Table \ref{Table_1_obs}, while the expected wage has small effects on the deviation of hours, consistent with expected hours responding to the expected wage almost as much as actual hours. As in the main estimates, relative losses and gains have similar effects.

\begin{table}[!htbp] \centering 
  \caption{Labor supply elasticity estimates, observational window, first differences} 
  \label{Table_dod_obs} 

\begin{adjustbox}{width=0.95\textwidth,center}

\def\sym#1{\ifmmode^{#1}\else\(^{#1}\)\fi}

\begin{tabular}{l*{6}{c}}
\hline\hline
                    &\multicolumn{1}{c}{(1)}&\multicolumn{1}{c}{(2)}&\multicolumn{1}{c}{(3)}&\multicolumn{1}{c}{(4)}&\multicolumn{1}{c}{(5)}&\multicolumn{1}{c}{(6)}\\
Effort:     &\multicolumn{6}{c}{Hours worked $-$ expected hours (ln)}\\ \cmidrule(lr){2-7}
Wage:       &\multicolumn{3}{c}{Per task, real}&\multicolumn{3}{c}{Hourly, real}\\ \cmidrule(lr){2-4}\cmidrule(lr){5-7}
            & OLS & IV & IV & OLS & IV & IV \\
\hline
\multicolumn{7}{l}{\textit{Panel A: pooled}}\\
$\Delta$ Wage from expectations (ln)&      -0.012        &       0.106        &       0.054        &      -0.260\sym{***}&       0.047        &       0.045        \\
                    &     (0.035)        &     (0.055)        &     (0.060)        &     (0.021)        &     (0.026)        &     (0.025)        \\
Expected wage (ln)  &      -0.008        &       0.052        &       0.208\sym{**}&       0.065\sym{**}&       0.011        &       0.127\sym{*} \\
                    &     (0.037)        &     (0.075)        &     (0.070)        &     (0.023)        &     (0.068)        &     (0.061)        \\
\hline
\multicolumn{7}{l}{\textit{Panel B: $\Delta$ wage by sign of the deviation}}\\
Loss ($\Delta < 0$) &      -0.015        &       0.085        &       0.140        &      -0.167\sym{***}&       0.047        &       0.015        \\
                    &     (0.059)        &     (0.111)        &     (0.143)        &     (0.032)        &     (0.033)        &     (0.043)        \\
Gain ($\Delta > 0$) &      -0.004        &       0.130        &      -0.128        &      -0.442\sym{***}&       0.049        &       0.144        \\
                    &     (0.058)        &     (0.155)        &     (0.222)        &     (0.047)        &     (0.074)        &     (0.111)        \\
p-value, gain = loss&       0.906        &       0.851        &       0.437        &       0.000        &       0.981        &       0.373        \\
Observations with loss&         676        &         676        &         674        &         768        &         768        &         765        \\
Observations with gain&         643        &         643        &         641        &         556        &         556        &         552        \\
\hline
Instrument, actual wage:&          --        &Platform tasks        &Platform tasks        &          --        & Screen time        & Screen time        \\
Instrument, expected wage:&          --        &  Avg. tasks        &   Composite        &          --        & Screen time        &   Composite        \\ \hline
Time-Region F.E.    &         Yes        &         Yes        &         Yes        &         Yes        &         Yes        &         Yes        \\
Individual controls &         Yes        &         Yes        &         Yes        &         Yes        &         Yes        &         Yes        \\
SW F-test ($\Delta$ wage)&                    &     217.317        &     194.294        &                    &     248.226        &     442.290        \\
SW F-test (expected wage)&                    &     234.880        &     203.808        &                    &     122.386        &     132.950        \\
Adjusted R-Squared  &       0.059        &       0.004        &      -0.010        &       0.303        &      -0.067        &       0.011        \\
Observations        &        1327        &        1327        &        1323        &        1327        &        1327        &        1320        \\
\hline\hline

\end{tabular}

\end{adjustbox}
 \begin{tabular}{l}
\multicolumn{1}{p{0.95\textwidth}}{\footnotesize Notes: SE clustered by occupation-sector in parentheses. Regressors and instruments as in Table \ref{Table_1_obs}; the outcome is the log difference between hours worked and expected hours.}\\
\multicolumn{1}{l}{\footnotesize *p<.05, **p<.01, ***p<.001}\\
\end{tabular}

\end{table}

\FloatBarrier

\section{Reference points in the experimental window} \label{A:dynamic}

As workers may update their reference point once the experiment starts, we replace the stated expectation with the pay offered in a previous sequence $s'$:
\begin{equation}
\label{eq:exp_seq}
    \ln(h_{is}) = \alpha + \Delta_S \ln(w_{is}) \, \eta_I + \ln(w_{is'}) \, \eta_U + X_i^{'} \delta + \epsilon_{is}
\end{equation}

\noindent where $\Delta_S \ln(w_{is}) = \ln(w_{is}) - \ln(w_{is'})$ is the change in pay between sequence $s$ and a previous sequence $s'$, comparing the second with the first, the third with the second, and the third with the first. The change in pay is randomized, so that $\eta_I$ is identified without the need for instruments.

Table \ref{Table_2_exp_seq} reports the estimates. The intertemporal elasticity is 0.85 between the second and the first sequence, and 1.11 and 1.13 in the third sequence, relative to the second and the first one, while the uncompensated elasticity to the pay of the earlier sequence is around 0.5. A smaller elasticity in the first transition is consistent with workers not yet knowing that pay varies randomly across sequences: a change in pay may then be read as informative about the pay of later sequences, and thus partly as an anticipated change, so that the response lies between the uncompensated and the intertemporal elasticity. By the third sequence, changes in pay are treated as transitory, and the response is fully intertemporal. Losses and gains have similar effects between consecutive sequences (columns 1 and 2), while losses relative to the first sequence dominate responses in the third one (1.51 against 0.75 for gains, column 3, with a p-value of 0.035 for their difference). Results are similar in first differences, with the change in the time spent between the two sequences as the outcome (Table \ref{Table_dod_exp}).

\begin{table}[!htbp] \centering 
  \caption{Labor supply elasticity estimates, experimental window, earlier sequence as reference} 
  \label{Table_2_exp_seq} 

\begin{adjustbox}{width=0.95\textwidth,center}

\def\sym#1{\ifmmode^{#1}\else\(^{#1}\)\fi}

\begin{tabular}{l*{3}{c}}
\hline\hline
                    &\multicolumn{1}{c}{(1)}&\multicolumn{1}{c}{(2)}&\multicolumn{1}{c}{(3)}\\
Effort:      &Seq. 2 (ln)&Seq. 3 (ln)&Seq. 3 (ln)\\ \cmidrule(lr){2-2}\cmidrule(lr){3-3}\cmidrule(lr){4-4}
Reference:   &Seq. 1 wage&Seq. 2 wage&Seq. 1 wage\\ \cmidrule(lr){2-2}\cmidrule(lr){3-3}\cmidrule(lr){4-4}
             & OLS & OLS & OLS \\
\hline
\multicolumn{4}{l}{\textit{Panel A: pooled}}\\
$\Delta$ Wage from the earlier sequence (ln): $\eta_I$&       0.846\sym{***}&       1.110\sym{***}&       1.130\sym{***}\\
                    &     (0.114)        &     (0.122)        &     (0.107)        \\
Wage in the earlier sequence (ln): $\eta_U$&       0.463\sym{***}&       0.491\sym{***}&       0.510\sym{***}\\
                    &     (0.077)        &     (0.087)        &     (0.091)        \\
\hline
\multicolumn{4}{l}{\textit{Panel B: $\Delta$ wage by sign of the deviation}}\\
Loss ($\Delta < 0$) &       1.120\sym{***}&       1.161\sym{***}&       1.509\sym{***}\\
                    &     (0.170)        &     (0.234)        &     (0.207)        \\
Gain ($\Delta > 0$) &       0.587\sym{**}&       1.056\sym{***}&       0.749\sym{***}\\
                    &     (0.218)        &     (0.242)        &     (0.212)        \\
p-value, gain = loss&       0.096        &       0.798        &       0.035        \\
Observations with loss&         752        &         757        &         751        \\
Observations with gain&         749        &         741        &         734        \\
\hline
Time-Region F.E.    &         Yes        &         Yes        &         Yes        \\
Individual controls &         Yes        &         Yes        &         Yes        \\
Adjusted R-Squared  &       0.128        &       0.157        &       0.157        \\
Observations        &        1547        &        1542        &        1542        \\
\hline\hline

\end{tabular}

\end{adjustbox}
 \begin{tabular}{l}
\multicolumn{1}{p{0.95\textwidth}}{\footnotesize Notes: SE clustered by respondent and occupation-sector in parentheses. The reference is the bonus of an earlier sequence; the deviation is the randomized change in pay between the two sequences. Panel B: separate slopes for negative and positive deviations.}\\
\multicolumn{1}{l}{\footnotesize *p<.05, **p<.01, ***p<.001}\\
\end{tabular}

\end{table}

\begin{table}[!htbp] \centering 
  \caption{Labor supply elasticity estimates, experimental window, first differences} 
  \label{Table_dod_exp} 

\begin{adjustbox}{width=0.95\textwidth,center}

\def\sym#1{\ifmmode^{#1}\else\(^{#1}\)\fi}

\begin{tabular}{l*{3}{c}}
\hline\hline
                    &\multicolumn{1}{c}{(1)}&\multicolumn{1}{c}{(2)}&\multicolumn{1}{c}{(3)}\\
Effort:      & Seq. 2 $-$ seq. 1 & Seq. 3 $-$ seq. 2 & Seq. 3 $-$ seq. 1 \\
             & OLS & OLS & OLS \\
\hline
\multicolumn{4}{l}{\textit{Panel A: pooled}}\\
$\Delta$ Wage between sequences (ln)&       0.780\sym{***}&       1.481\sym{***}&       1.298\sym{***}\\
                    &     (0.096)        &     (0.094)        &     (0.122)        \\
Wage in the earlier sequence (ln)&       0.136        &       0.052        &       0.291\sym{*} \\
                    &     (0.098)        &     (0.063)        &     (0.113)        \\
\hline
\multicolumn{4}{l}{\textit{Panel B: $\Delta$ wage by sign of the deviation}}\\
Loss ($\Delta < 0$) &       0.749\sym{***}&       1.575\sym{***}&       1.543\sym{***}\\
                    &     (0.219)        &     (0.231)        &     (0.236)        \\
Gain ($\Delta > 0$) &       0.809\sym{***}&       1.381\sym{***}&       1.048\sym{***}\\
                    &     (0.188)        &     (0.212)        &     (0.247)        \\
p-value, gain = loss&       0.865        &       0.629        &       0.237        \\
Observations with loss&         620        &         638        &         629        \\
Observations with gain&         629        &         609        &         606        \\
\hline
Time-Region F.E.    &         Yes        &         Yes        &         Yes        \\
Individual controls &         Yes        &         Yes        &         Yes        \\
Adjusted R-Squared  &       0.132        &       0.259        &       0.165        \\
Observations        &        1284        &        1279        &        1279        \\
\hline\hline

\end{tabular}

\end{adjustbox}
 \begin{tabular}{l}
\multicolumn{1}{p{0.95\textwidth}}{\footnotesize Notes: SE clustered by respondent and occupation-sector in parentheses. One observation per respondent; the outcome is the log difference between the time spent in the later and in the earlier sequence, and the change in pay between the two sequences is randomized. Workers whose target pay was capped are excluded.}\\
\multicolumn{1}{l}{\footnotesize *p<.05, **p<.01, ***p<.001}\\
\end{tabular}

\end{table}

Table \ref{Table_dynamic} examines which reference point workers use in the experiment. Column 1 splits the deviation from the target between the first sequence and the later ones. The response is much smaller in the first sequence (0.35) than in the second and third (1.49), and the difference is statistically significant at the 0.1\% level. The difference does not depend on whether the target pay has already been offered in an earlier sequence (column 2, 1.56 and 1.31, with a p-value of 0.28 for their difference). The response increases with the time spent in the earlier sequences (0.43 for one standard deviation), and does not depend on the share of correct grids in the previous sequence (column 8), so that it grows with exposure to the task rather than declining with fatigue.

Columns 3 to 6 compare the stated target with the pay of the previous sequence as reference points. Each of the two deviations affects effort when it enters the regression alone (columns 3 and 4), but when both enter the same regression the deviation from the target keeps its effect (1.45), while the change from the previous sequence has none (0.03, column 5). The result is unchanged when the target pay is instrumented with the random anchor (1.66 and $-0.08$, column 6). When the pay of the previous sequence is the reference, and is instrumented with the random anchor and with the randomized factor of the previous sequence (column 7), its coefficient (1.48) equals that of the change in pay (1.49), so that effort only depends on the pay of the current sequence. Workers therefore do not update their reference point to the pay received in earlier sequences, and keep responding to deviations from the expectation stated before the experiment.

\begin{sidewaystable}[!htbp] \centering 
  \caption{Reference points in the experimental window} 
  \label{Table_dynamic} 

\begin{adjustbox}{width=0.95\textwidth,center}

\def\sym#1{\ifmmode^{#1}\else\(^{#1}\)\fi}

\begin{tabular}{l*{8}{c}}
\hline\hline
                    &\multicolumn{1}{c}{(1)}&\multicolumn{1}{c}{(2)}&\multicolumn{1}{c}{(3)}&\multicolumn{1}{c}{(4)}&\multicolumn{1}{c}{(5)}&\multicolumn{1}{c}{(6)}&\multicolumn{1}{c}{(7)}&\multicolumn{1}{c}{(8)}\\
Effort:      &\multicolumn{8}{c}{Hours in the sequence (ln)}\\ \cmidrule(lr){2-9}
Sequences:   & All & \multicolumn{7}{c}{Second and third}\\ \cmidrule(lr){2-2}\cmidrule(lr){3-9}
             & OLS & OLS & OLS & OLS & OLS & IV & IV & OLS \\
\hline
$\Delta$ from the target (ln)&                    &                    &       1.490\sym{***}&                    &       1.446\sym{***}&       1.660\sym{***}&                    &       1.644\sym{***}\\
                    &                    &                    &     (0.096)        &                    &     (0.213)        &     (0.274)        &                    &     (0.088)        \\
\quad first sequence&       0.345\sym{**}&                    &                    &                    &                    &                    &                    &                    \\
                    &     (0.131)        &                    &                    &                    &                    &                    &                    &                    \\
\quad second and third sequence&       1.487\sym{***}&                    &                    &                    &                    &                    &                    &                    \\
                    &     (0.097)        &                    &                    &                    &                    &                    &                    &                    \\
\quad target pay already offered&                    &       1.562\sym{***}&                    &                    &                    &                    &                    &                    \\
                    &                    &     (0.113)        &                    &                    &                    &                    &                    &                    \\
\quad target pay not yet offered&                    &       1.306\sym{***}&                    &                    &                    &                    &                    &                    \\
                    &                    &     (0.205)        &                    &                    &                    &                    &                    &                    \\
\quad $\times$ time in earlier sequences (std.)&                    &                    &                    &                    &                    &                    &                    &       0.425\sym{***}\\
                    &                    &                    &                    &                    &                    &                    &                    &     (0.105)        \\
\quad $\times$ correct grids, previous sequence (std.)&                    &                    &                    &                    &                    &                    &                    &      -0.038        \\
                    &                    &                    &                    &                    &                    &                    &                    &     (0.116)        \\
$\Delta$ from the previous sequence (ln)&                    &                    &                    &       1.009\sym{***}&       0.030        &      -0.078        &       1.492\sym{***}&                    \\
                    &                    &                    &                    &     (0.076)        &     (0.129)        &     (0.161)        &     (0.145)        &                    \\
Target pay (ln)     &       0.417\sym{***}&       0.464\sym{***}&       0.466\sym{***}&                    &       0.464\sym{***}&       1.011\sym{*} &                    &       0.284\sym{***}\\
                    &     (0.070)        &     (0.087)        &     (0.087)        &                    &     (0.087)        &     (0.401)        &                    &     (0.066)        \\
Pay in the previous sequence (ln)&                    &                    &                    &       0.512\sym{***}&                    &                    &       1.481\sym{***}&                    \\
                    &                    &                    &                    &     (0.083)        &                    &                    &     (0.290)        &                    \\
Time in earlier sequences (std.)&                    &                    &                    &                    &                    &                    &                    &       0.759\sym{***}\\
                    &                    &                    &                    &                    &                    &                    &                    &     (0.030)        \\
Correct grids, previous sequence (std.)&                    &                    &                    &                    &                    &                    &                    &       0.053        \\
                    &                    &                    &                    &                    &                    &                    &                    &     (0.038)        \\
p-value, equal slopes&       0.000        &       0.284        &                    &                    &                    &                    &                    &                    \\
\hline
Instrument:         &          --        &          --        &          --        &          --        &          --        &      Anchor        &Anchor, previous factor        &          --        \\ \hline
Sequence F.E.       &         Yes        &         Yes        &         Yes        &         Yes        &         Yes        &         Yes        &         Yes        &         Yes        \\
Time-Region F.E.    &         Yes        &         Yes        &         Yes        &         Yes        &         Yes        &         Yes        &         Yes        &         Yes        \\
Individual controls &         Yes        &         Yes        &         Yes        &         Yes        &         Yes        &         Yes        &         Yes        &         Yes        \\
SW F-test (level)   &                    &                    &                    &                    &                    &      76.971        &      86.430        &                    \\
Adjusted R-Squared  &       0.199        &       0.151        &       0.151        &       0.145        &       0.151        &       0.072        &       0.010        &       0.365        \\
Observations        &        3861        &        2565        &        2565        &        2565        &        2565        &        2565        &        2565        &        2470        \\
\hline\hline

\end{tabular}

\end{adjustbox}
 \begin{tabular}{l}
\multicolumn{1}{p{0.95\textwidth}}{\footnotesize Notes: SE clustered by respondent and occupation-sector in parentheses. Workers whose target pay was set at the maximum of 0.22 pounds per task are excluded. $\Delta$ from the target: log of the randomized factor applied to the target pay. $\Delta$ from the previous sequence: log change in pay with respect to the previous sequence. Column 1: deviation from the target in the first and in later sequences. Column 2: deviation from the target by whether the target pay was offered in an earlier sequence. Columns 5-6: both deviations, with the target pay instrumented with the random anchor in column 6. Column 7: the pay in the previous sequence is instrumented with the random anchor and with the randomized factor of the previous sequence. Column 8: deviation from the target interacted with the (log) time spent in the earlier sequences and with the (log) share of correct grids in the previous sequence, both standardized.}\\
\multicolumn{1}{l}{\footnotesize *p<.05, **p<.01, ***p<.001}\\
\end{tabular}

\end{sidewaystable}

\FloatBarrier

\section{Comparison between windows with the single instrument} \label{A:single}

Tables \ref{Table_3_windows_single} and \ref{Table_recall_single} replicate Tables \ref{Table_3_windows} and \ref{Table_recall}, with the observational expected wage instrumented with the single instrument of Table \ref{Table_1_obs} (columns 2 and 5) rather than with the composite instrument. In Table \ref{Table_3_windows_single}, the observational uncompensated elasticities are larger (1.20 and 1.33 without individual fixed effects, 1.00 and 1.52 with them), and the observational intertemporal elasticities are smaller (between 0.15 and 0.29), so that the observational intertemporal elasticity is between 0.78 and 0.89 lower than the experimental one. The observational uncompensated elasticity remains statistically indistinguishable from the experimental intertemporal elasticity in all columns. In Table \ref{Table_recall_single}, the observational intertemporal elasticity of a worker with perfect recall of their earnings is between 0.55 and 0.79, and decreases with the recall error, while its difference from the experimental one is between 0.32 and 0.58, statistically significant only for the hourly wage without individual fixed effects. With the single instrument, the observational uncompensated elasticity also decreases with the recall error.

\begin{table}[!htbp] \centering
  \caption{Labor supply elasticity estimates, observational and experimental windows within respondents (single instrument)}
  \label{Table_3_windows_single}
\def\sym#1{\ifmmode^{#1}\else\(^{#1}\)\fi}
\small
\begin{tabular}{lcccc}
\hline\hline
 & (1) & (2) & (3) & (4) \\
Effort: & \multicolumn{4}{c}{Hours of work (ln)} \\ \cmidrule(lr){2-5}
Observational wage: & \multicolumn{2}{c}{Per task, real} & \multicolumn{2}{c}{Hourly, real} \\ \cmidrule(lr){2-3}\cmidrule(lr){4-5}
Individual F.E.: & No & Yes & No & Yes \\
\hline
\multicolumn{5}{l}{\textit{Panel A: $\Delta$ Wage from expectations ($\eta_I$)}} \\
\quad Observational & 0.291\sym{**} & 0.145 & 0.176\sym{*} & 0.177 \\
 & (0.098) & (0.134) & (0.076) & (0.094) \\
\quad Experimental & 1.067\sym{***} & 1.032\sym{***} & 1.067\sym{***} & 1.035\sym{***} \\
 & (0.075) & (0.068) & (0.075) & (0.070) \\
\quad p-value, observational = experimental & 0.000 & 0.000 & 0.000 & 0.000 \\
\addlinespace
\multicolumn{5}{l}{\textit{Panel B: Expected wage ($\eta_U$)}} \\
\quad Observational & 1.202\sym{***} & 0.997\sym{***} & 1.332\sym{***} & 1.520\sym{***} \\
 & (0.131) & (0.187) & (0.195) & (0.288) \\
\quad Experimental & 0.812\sym{**} & 0.487\sym{***} & 0.816\sym{**} & 0.730\sym{***} \\
 & (0.265) & (0.105) & (0.266) & (0.160) \\
\quad p-value, observational = experimental & 0.157 & 0.003 & 0.083 & 0.002 \\
\addlinespace
\multicolumn{5}{l}{\textit{Panel C: $\Delta$ Wage, losses ($\Delta < 0$)}} \\
\quad Observational & 0.190 & 0.049 & 0.144 & 0.081 \\
 & (0.201) & (0.270) & (0.090) & (0.112) \\
\quad Experimental & 1.406\sym{***} & 1.361\sym{***} & 1.378\sym{***} & 1.385\sym{***} \\
 & (0.164) & (0.147) & (0.163) & (0.163) \\
\quad p-value, observational = experimental & 0.000 & 0.000 & 0.000 & 0.000 \\
\addlinespace
\multicolumn{5}{l}{\textit{Panel D: $\Delta$ Wage, gains ($\Delta > 0$)}} \\
\quad Observational & 0.426 & 0.304 & 0.245 & 0.435 \\
 & (0.262) & (0.391) & (0.198) & (0.268) \\
\quad Experimental & 0.706\sym{***} & 0.693\sym{***} & 0.713\sym{***} & 0.650\sym{***} \\
 & (0.187) & (0.150) & (0.187) & (0.156) \\
\quad p-value, observational = experimental & 0.378 & 0.368 & 0.064 & 0.478 \\
\addlinespace
\multicolumn{5}{l}{\textit{Panel E: Tests between margins}} \\
\quad p-value, observational $\eta_I = \eta_U$ & 0.000 & 0.000 & 0.000 & 0.000 \\
\quad p-value, experimental $\eta_I = \eta_U$ & 0.308 & 0.000 & 0.319 & 0.065 \\
\quad p-value, observational $\eta_U$ = experimental $\eta_I$ & 0.388 & 0.861 & 0.206 & 0.103 \\
\addlinespace Observations (respondents), Panels A-B, E & 4406 (1097) & 4432 (1101) & 4397 (1094) & 4420 (1098) \\
Observations (respondents), Panels C-D & 4382 (1102) & 4408 (1095) & 4385 (1102) & 4408 (1095) \\
\hline\hline
\end{tabular}
 \begin{tabular}{l}
\multicolumn{1}{p{0.95\textwidth}}{\footnotesize Notes: As Table \ref{Table_3_windows}, with the observational expected wage instrumented with the single instrument of Table \ref{Table_1_obs} (columns 2 and 5). }\\
\multicolumn{1}{l}{\footnotesize *p<.05, **p<.01, ***p<.001}\\
\end{tabular}
\end{table}

\begin{sidewaystable}[!htbp] \centering
  \caption{Labor supply elasticity estimates by recall accuracy, observational and experimental windows (single instrument)}
  \label{Table_recall_single}
\def\sym#1{\ifmmode^{#1}\else\(^{#1}\)\fi}
\begin{adjustbox}{width=0.95\textwidth,center}
\begin{tabular}{lcccccccc}
\hline\hline
 & (1) & (2) & (3) & (4) & (5) & (6) & (7) & (8) \\
Effort: & \multicolumn{8}{c}{Hours of work (ln)} \\ \cmidrule(lr){2-9}
Recall error: & \multicolumn{4}{c}{Earnings} & \multicolumn{4}{c}{Tasks} \\ \cmidrule(lr){2-5}\cmidrule(lr){6-9}
Observational wage: & \multicolumn{2}{c}{Per task, real} & \multicolumn{2}{c}{Hourly, real} & \multicolumn{2}{c}{Per task, real} & \multicolumn{2}{c}{Hourly, real} \\ \cmidrule(lr){2-3}\cmidrule(lr){4-5}\cmidrule(lr){6-7}\cmidrule(lr){8-9}
Individual F.E.: & No & Yes & No & Yes & No & Yes & No & Yes \\
\hline
\quad \textit{$\Delta$ Wage from expectations ($\eta_I$)} \\
\quad \quad Observational & 0.719\sym{***} & 0.791\sym{**} & 0.548\sym{**} & 0.618\sym{*} & 0.554\sym{***} & 0.421\sym{*} & 0.331\sym{*} & 0.349 \\
 & (0.205) & (0.244) & (0.202) & (0.244) & (0.156) & (0.166) & (0.147) & (0.182) \\
\quad \quad \quad $\times$ recall error & -0.303\sym{**} & -0.462\sym{***} & -0.219\sym{**} & -0.301\sym{***} & -0.354\sym{*} & -0.406\sym{*} & -0.171\sym{*} & -0.210\sym{*} \\
 & (0.094) & (0.104) & (0.080) & (0.091) & (0.169) & (0.165) & (0.080) & (0.086) \\
\quad \quad Experimental & 1.125\sym{***} & 1.109\sym{***} & 1.124\sym{***} & 1.091\sym{***} & 1.080\sym{***} & 1.085\sym{***} & 1.078\sym{***} & 1.073\sym{***} \\
 & (0.111) & (0.101) & (0.112) & (0.103) & (0.119) & (0.108) & (0.119) & (0.109) \\
\quad \quad \quad $\times$ recall error & -0.079 & -0.103 & -0.078 & -0.085 & -0.016 & -0.082 & -0.015 & -0.060 \\
 & (0.104) & (0.096) & (0.105) & (0.099) & (0.130) & (0.121) & (0.130) & (0.125) \\
\quad \quad p-value, observational = experimental at zero recall error & 0.080 & 0.257 & 0.013 & 0.091 & 0.009 & 0.001 & 0.000 & 0.001 \\
\quad \textit{Expected wage ($\eta_U$)} \\
\quad \quad Observational & 1.584\sym{***} & 1.269\sym{***} & 2.028\sym{***} & 2.500\sym{***} & 1.353\sym{***} & 1.199\sym{***} & 1.685\sym{***} & 2.144\sym{***} \\
 & (0.222) & (0.244) & (0.506) & (0.667) & (0.218) & (0.259) & (0.388) & (0.567) \\
\quad \quad \quad $\times$ recall error & -0.353\sym{**} & -0.310\sym{*} & -0.497 & -0.768\sym{*} & -0.249 & -0.334 & -0.399 & -0.707\sym{*} \\
 & (0.120) & (0.144) & (0.255) & (0.299) & (0.190) & (0.258) & (0.231) & (0.309) \\
\quad \quad Experimental & 0.745\sym{*} & 0.540\sym{***} & 0.753\sym{*} & 0.817\sym{**} & 0.667 & 0.649\sym{***} & 0.673 & 0.773\sym{**} \\
 & (0.318) & (0.145) & (0.319) & (0.298) & (0.378) & (0.146) & (0.379) & (0.252) \\
\quad \quad \quad $\times$ recall error & 0.067 & -0.100 & 0.064 & -0.091 & 0.205 & -0.238 & 0.202 & -0.078 \\
 & (0.230) & (0.118) & (0.230) & (0.167) & (0.305) & (0.156) & (0.305) & (0.213) \\
\quad \quad p-value, observational = experimental at zero recall error & 0.037 & 0.002 & 0.028 & 0.002 & 0.105 & 0.038 & 0.049 & 0.005 \\
\addlinespace Observations (respondents) & 4398 (1106) & 4424 (1099) & 4389 (1103) & 4412 (1096) & 4406 (1108) & 4432 (1101) & 4397 (1105) & 4420 (1098) \\
\hline\hline
\end{tabular}
\end{adjustbox}
 \begin{tabular}{l}
\multicolumn{1}{p{0.95\textwidth}}{\footnotesize Notes: As Table \ref{Table_recall}, with the observational expected wage instrumented with the single instrument of Table \ref{Table_1_obs} (columns 2 and 5).}\\
\multicolumn{1}{l}{\footnotesize *p<.05, **p<.01, ***p<.001}\\
\end{tabular}
\end{sidewaystable}

\FloatBarrier

\section{Selection of the moderators} \label{A:lasso}

The candidate moderators are all the worker characteristics we observe: the individual controls, weekly targets for earnings, hours, and tasks, the share of time spent searching for studies, the screen time of the other weeks, and the recall errors in earnings and tasks. They are standardized (or centered, if binary), so that the main terms are the elasticities of a worker with average characteristics. Each characteristic is interacted with the four wage terms of Equation \ref{eq:within} and with the indicator of the reference week. After partialling out individual and period fixed effects and the wage terms, with the observational wage terms replaced by their first-stage predictions, the interactions are selected with the heteroskedasticity-robust plug-in penalty of \citet{belloni2012sparse} and, as an alternative, by cross-validation with folds by worker. Equation \ref{eq:within} is then estimated with all the interactions of every characteristic selected at least once.

Table \ref{Table_lasso} reports the selected interactions and the post-lasso estimates. The recall error in earnings is the only characteristic selected as a moderator of the observational intertemporal elasticity. In the post-lasso estimates, one standard deviation of recall error lowers the observational intertemporal elasticity by 0.46 (per-task wage) and 0.32 (hourly wage), while it has no statistically significant effect on the observational uncompensated elasticity ($-0.18$ and $-0.25$) and leaves the experimental elasticities unchanged. The other selected characteristics, gender and the share of time spent searching for studies, have no statistically significant interaction with any of the elasticities.

\begin{table}[!htbp] \centering
  \caption{Moderators of the elasticities selected by lasso}
  \label{Table_lasso}
\def\sym#1{\ifmmode^{#1}\else\(^{#1}\)\fi}
\small
\begin{tabular}{lcc}
\hline\hline
 & (1) & (2) \\
Observational wage: & Per task, real & Hourly, real \\
\hline
\multicolumn{3}{l}{\textit{Elasticities of the average worker}} \\
\quad Intertemporal, observational & 0.478\sym{*} & 0.515\sym{***} \\
 & (0.207) & (0.147) \\
\quad Uncompensated, observational & 0.859\sym{***} & 1.113\sym{***} \\
 & (0.178) & (0.242) \\
\quad Intertemporal, experimental & 1.033\sym{***} & 1.040\sym{***} \\
 & (0.066) & (0.063) \\
\quad Uncompensated, experimental & 0.427\sym{***} & 0.514\sym{***} \\
 & (0.105) & (0.131) \\
\addlinespace
\multicolumn{3}{l}{\textit{$\times$ Sex}} \\
\quad Intertemporal, observational & -0.281 & -0.018 \\
 & (0.227) & (0.135) \\
\quad Uncompensated, observational & 0.369 & 0.333 \\
 & (0.246) & (0.316) \\
\quad Intertemporal, experimental & -0.120 & -0.125 \\
 & (0.146) & (0.149) \\
\quad Uncompensated, experimental & 0.219 & 0.363 \\
 & (0.183) & (0.266) \\
\addlinespace
\multicolumn{3}{l}{\textit{$\times$ Share of time searching for studies}} \\
\quad Intertemporal, observational & 0.092 & 0.063 \\
 & (0.148) & (0.071) \\
\quad Uncompensated, observational & -0.107 & -0.212 \\
 & (0.169) & (0.179) \\
\quad Intertemporal, experimental & -0.064 & -0.050 \\
 & (0.075) & (0.070) \\
\quad Uncompensated, experimental & -0.057 & -0.038 \\
 & (0.098) & (0.130) \\
\addlinespace
\multicolumn{3}{l}{\textit{$\times$ Recall error, earnings}} \\
\quad Intertemporal, observational & -0.460\sym{***} & -0.322\sym{***} \\
 & (0.093) & (0.064) \\
\quad Uncompensated, observational & -0.177 & -0.251 \\
 & (0.130) & (0.148) \\
\quad Intertemporal, experimental & -0.086 & -0.075 \\
 & (0.076) & (0.077) \\
\quad Uncompensated, experimental & -0.083 & -0.006 \\
 & (0.095) & (0.114) \\
\hline\hline
\end{tabular}
 \begin{tabular}{l}
\multicolumn{1}{p{0.95\textwidth}}{\footnotesize Notes: SE clustered by occupation-sector in parentheses. IV estimates of the model of Table \ref{Table_3_windows} with individual F.E., with the wage terms interacted with the characteristics selected by lasso among 120 candidate interactions (Section \ref{s:framework}), standardized (or centred, if binary), so that the main terms are the elasticities of the average worker. Selected interactions, Per task, real: plug-in penalty, Sex $\times$ reference week; Share of time searching for studies $\times$ experimental expected wage; Share of time searching for studies $\times$ reference week; Recall error, earnings $\times$ observational $\Delta$; cross-validation, none. Hourly, real: plug-in penalty, Share of time searching for studies $\times$ reference week; Recall error, earnings $\times$ observational $\Delta$; cross-validation, Share of time searching for studies $\times$ reference week; Recall error, earnings $\times$ observational $\Delta$.}\\
\multicolumn{1}{l}{\footnotesize *p<.05, **p<.01, ***p<.001}\\
\end{tabular}
\end{table}

\FloatBarrier

\FloatBarrier

\end{document}